\documentclass[sigconf,10pt]{acmart}

\usepackage{tikz}
\usepackage{amsmath}
\usepackage{amsthm}

\usepackage{preamble}
\usepackage{aliases}
\usepackage{appendix}
\usepackage[normalem]{ulem}

\renewcommand\footnotetextcopyrightpermission[1]{}
\begin{document}

\date{}

\title[\name]{\name: An Open Research Framework for Optical Data Center Networks
}

\author{Yiming Lei}
\affiliation{
  \institution{Max Planck Institute for Informatics}
  \country{}
}

\author{Federico De Marchi}
\affiliation{
  \institution{Max Planck Institute for Informatics}
  \country{}
}

\author{Jialong Li}
\affiliation{
  \institution{Max Planck Institute for Informatics}
  \country{}
}

\author{Raj Joshi}
\affiliation{
  \institution{Harvard University}
  \country{}
}

\author{Balakrishnan Chandrasekaran}
\affiliation{
  \institution{Vrije Universiteit Amsterdam}
  \country{}
}

\author{Yiting Xia}
\affiliation{
  \institution{Max Planck Institute for Informatics}
  \country{}
}

\begin{abstract}

Optical data center networks (DCNs) are emerging as a promising design for cloud infrastructure. However, existing optical DCN architectures operate as closed ecosystems, tying software solutions to specific optical hardware. We introduce \name, an open research framework that decouples software from hardware, allowing them to evolve independently.
\name{} features: (1) a time-flow table abstraction as a common interface between optical hardware and software, (2) a unified workflow and user-friendly API for implementing various optical \dcns{} with simple Python scripts, and (3) a backend system that re-architects queue management to support the time-flow tables and provides rich infrastructure services for diverse applications.
Built on programmable switches, \name{} achieves a record-breaking minimum optical circuit duration of \umus{2} using commodity devices. We validate \name{}’s generality by implementing six optical architectures and seven routing schemes on an optical testbed and conducting benchmarks on a 108-ToR setup, showcasing its efficiency. Additionally, case studies highlight novel research opportunities enabled by \name{}.

\end{abstract}

\maketitle

\section{Introduction}\label{sec:intro}

In the post-Moore’s law era for merchant silicon, the networking community has turned to optical circuit switch (OCS) technologies for their power, cost, and bandwidth advantages. Numerous optical data center network (\dcn) architectures have been proposed to build high-performance \dcn fabrics with different \textit{\ocs hardware}~\cite{Helios, cThrough, OSA, Mordia, Quartz, MegaSwitch, WaveCube, ProjectToR, Flat-tree, ShareBackup, OmniSwitch, RotorNet, Sirius, Opera, SiP-ML, TopoOpt, JupiterEvolving, Shale, NegotiaToR, RealizeRotorNet}.

\begin{figure}[tb]
    \centering%
    \includegraphics[width=\linewidth]
    {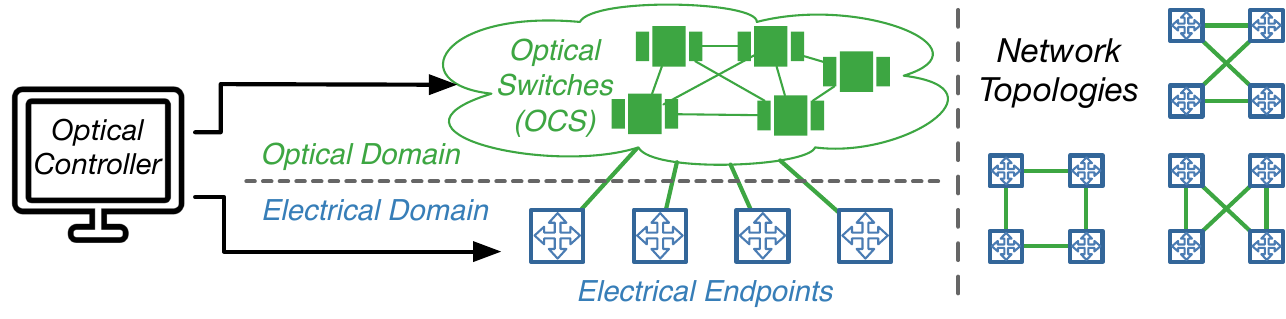}
    \figcap{An example optical \dcn and its three topologies connected by different optical circuits over the \ocses.}
    \label{fig:optical-dcn-example}
\end{figure}

\ocses are, however, bufferless physical-layer devices that transmit only optical signals. They establish exclusive optical circuits between electrical communication endpoints, such as pods, Top-of-Rack switches (ToRs), and host NICs, and reconfigure the circuits to form varying network topologies (Fig.~\ref{fig:optical-dcn-example}). Hence, to make an optical DCN functional, a \textit{software system} must coordinate packet buffering and scheduling on these endpoints to align with the circuit availability.

The \textit{tight coupling of optical hardware and software systems} makes research and innovation for optical \dcns siloed. Each optical \dcn architecture is a \textit{closed ecosystem} comprising specialized optical hardware and customized networked systems to support that hardware.
System solutions are tied to the underlying hardware architecture, creating a high \textit{barrier to entry} for system researchers with little optics expertise.

Equally limited is the \textit{experimental environment}. The few evaluation platforms for optical \dcns emulate switches on hosts without the actual switch stack, achieving only modest scale (8 ToRs) and capacity (\uGbps{10})~\cite{reTCP, ExRec}. Widely used network simulators~\cite{NS3, OMNeT, DONS, OPNET} lack native support for optical \dcns and, even if extended, are confined to packet- or flow-level analysis, thus inadequate for the broad exploration needs of this actively studied field.
The lack of realistic and general experimental environment forces system designs to rely on home-grown simulators and minimal-scale testbeds specific to the optical architecture, making it difficult to validate, compare, and improve different proposals, further stalling the innovation cycle.

In this paper, we present \name, an open research framework to \textit{decouple software systems from optical hardware, enabling them to evolve independently.} We aim to make \name a ``narrow waist'' for optical \dcns, building a general system stack
that supports easy integration of various optical hardware downward and free exploration of network protocols upward.
As a collaborative research sandbox, \name accelerates problem exposure and resolution across the entire network stack, paving the way for the deployment of diverse optical \dcn solutions in practice.

We addressed a series of design and engineering challenges achieving this ambitious goal, making both methodological and technical contributions to the state of the art.

The primary challenge is defining a \textit{common interface} between hardware and software across various optical architectures. Like the IP layer in traditional networks, we believe this ``narrow waist'' lies in routing, as the core function of optical \dcns is to direct traffic through the dynamic circuits.
This interface must (1) capture the new requirement for routing packets across changing topologies; 
(2) resemble classical network abstractions to help system researchers adapt quickly to the new topic while hiding optical hardware details;
(3) be expressive enough to encompass existing and future routing schemes in optical \dcns; and (4) maintain backward compatibility with classical abstractions to support traditional \dcns for performance comparisons.

We propose the \textit{time-flow table} abstraction (\S\ref{sec:table}) to unify diverse optical \dcn architectures. This abstraction emphasizes the \textit{notion of time}, reflecting the \textit{circuit switching} nature of optical \dcns where circuits are discrete and persist for fixed time periods, despite variations in hardware and software designs. The time-flow table enables packets to exit a switch or NIC not immediately upon arrival, but at a scheduled time aligned with circuit availability. This abstraction can represent all routing primitives used in existing and potential future routing schemes for optical \dcns, and it reduces to a regular flow table for traditional \dcns when the time-related fields are set to wildcards.

The second challenge is defining a \textit{programming model} that simplifies the implementation of existing optical \dcn architectures and opens up new designs. Optical \dcns fall into two categories: \textit{traffic-aware} (\tra), which dynamically reconfigure the topology based on traffic patterns~\cite{Helios, cThrough, REACToR, ProjectToR, MegaSwitch, Mordia, NegotiaToR, JupiterEvolving, OSA, Flat-tree, WaveCube, Lightwave}, and \textit{traffic-oblivious} (\tro), which operate using a pre-determined circuit schedule independent of traffic conditions~\cite{Sirius, Shale, RotorNet, Opera, RealizeRotorNet}. These distinct workflows create a clear boundary between \tra and \tro designs, limiting exploration of the broader design space. 
Even within a design class, implementing an optical architecture remains labor-intensive, requiring meticulous management of individual circuits and routing updates as the topology reconfigures.

We unify the \tra and \tro \textit{workflows} and design a set of common \textit{API functions} (\S\ref{sec:api}) that allow users to program specific topologies and routing strategies with simple Python scripts. The workflow supports both online updates for \tra designs and offline pre-loading of deterministic configurations for \tro ones. These APIs define high-level network behaviors and compile them into low-level \ocs connections and time-flow table entries.
In $\S$\ref{sec:api}, we exemplify how these APIs realize existing \tra and \tro architectures and enable future hybrid designs, e.g., using \tro for connection-rich intra-rack scale-up networks and \tra for locality-friendly inter-rack scale-out networks for emerging machine learning workloads.

The third challenge lies in designing a \textit{backend system} to support the time-flow table. Rather than simply extending flow tables with additional fields, the time-flow table abstraction introduces a fundamental shift in packet processing. It requires time-based packet lookup, buffering, and scheduling, which go beyond traditional priority-based queue management and pose significant challenges for switches and NICs with limited buffers. Many proposals cannot be implemented end-to-end without such support~\cite{Sirius, Shale, RotorNet, HOHO, UCMP, NegotiaToR}, and the implemented ones~\cite{Helios, cThrough, OSA, JupiterEvolving, Mordia, Flat-tree, ProjectToR, RealizeRotorNet, Opera} have to rely on legacy flow tables for static topologies, restricting routing to individual topology instances or a subset of continuously available circuits on the changing topologies.

We leverage P4 packet processing programmability to re-architect the \textit{queue management system} (\S\ref{sec:q-rotation}).
By synchronizing switches and NICs with optical circuits at nanosecond precision~\cite{OpSync},\footnote{The synchronization protocol is described in a separate paper.} the system performs time-based packet lookup and queue allocation based on circuit availability, pausing and unpausing appropriate queues to buffer packets and release them precisely when circuits become available.

The backend system also provides rich \textit{infrastructure services} (\S\ref{sec:infra_services}) including common system requirements, such as ensuring scheduled data completes transmission within circuit durations under congestion, preventing losses in the optical fabric.
It, therefore, offers a foundation on top which users can investigate various upper-layer network protocol designs.  
We offer, for instance, congestion detection, but users define the (mitigation) responses, along with a traffic push-back mechanism as a last-resort flow control.
It also offers optimization ``knobs,'' as proposed in various optical architectures, but not implemented before.

We built \name using Intel Tofino2 switches and the \vma userspace library on Mellanox NICs.
The host-only version with Corundum FPGA-based NICs~\cite{Corundum} is under development. \name provides a 
\textit{full ecosystem} for users with different hardware availability.
Users with real \ocses can plug them into \name for an end-to-end functional system, requiring no changes to the system stack when upgrading \ocs hardware. For users without real \ocses, we provide an emulated optical network fabric using programmable switches (\S\ref{sec:mininet}), emulating different types and structures of \ocses.
Lastly, we offer a Mininet educational toolkit (\S\ref{sec:mininet}) that runs \name in a virtual network for those without access to programmable switches.

We evaluate \name on a testbed with real and emulated \ocses running cloud applications. To demonstrate its generality, we implement six \tra and \tro architectures and seven routing schemes (\S\ref{sec:research}). Though these prior work are either proprietary or only partially implemented, we validate \name{}'s correctness by inferring performance trends from prior work and reproducing limited results with comparable settings. We also showcase new research use cases \name enables (\S\ref{sec:research})
and conduct extensive benchmarking studies at scale (\S\ref{sec:perf}). In a 108-ToR setting with production \dcn traces, all system components function efficiently, keeping usage of switch buffer and other hardware resources under 13.8\%. \name supports a minimum circuit duration of \umus{2}---the lowest ever achieved with commodity network devices, offering guidance for practical optical \dcn deployment. We will open-source \name on acceptance to foster future research on the topic.

\textit{[This work does not raise any ethical issues.]}

\section{Background}\label{sec:background}

We briefly introduce optical \dcns and their routing strategies to reveal key requirements for the \name design.

\subsection{Optical \dcns}\label{sec:optical-dcns}

An optical \dcn has an \textit{optical network fabric} in the core, formed by interconnecting and configuring OCSes to realize a concerned DCN architecture (Fig.~\ref{fig:optical-dcn-example}).
\ocses transmit optical signals in the \textit{physical layer}, acting as waveguides (similar to an optical fiber), but with the additional capability of circuit reconfiguration.
An \textit{optical controller}, e.g., a server or FPGA board, controls the \ocses to establish \textit{optical circuits} between electrical communication endpoints, such as pods, ToRs, and host NICs,
where a pair of them has exclusive access to the circuit.
The controller reconfigures these circuits to create different network topologies (refer~Fig.~\ref{fig:optical-dcn-example}). As waveguides, \ocses are bufferless and cannot store or process packets, requiring the communication endpoints to be coordinated to send packets only when the optical circuits are available.
This coordination is usually achieved using the optical controller, although endpoints can also self-organize (i.e., coordinate amongst themselves).
Optical \dcns are broadly classified as either traffic-aware (\tra) or traffic-oblivious (\tro), based on how they reconfigure the circuits.

\textbf{\tra optical \dcns} establish circuits based on traffic demands, but the control loop, involving traffic collection and topology computation takes milliseconds to seconds, delaying latency-sensitive ``mice'' flows as they wait for circuits to be ready.
One approach (\textit{TA-1}), hence, uses a static network for mice flows, while sending elephant flows opportunistically over the optical circuits. The static network can either be a parallel electrical \dcn, creating a so called electrical-optical hybrid \dcn~\cite{Helios, cThrough, REACToR}, or a default topology with some stable circuits~\cite{ProjectToR, MegaSwitch}.
Another solution (\textit{TA-2}) ensures every reconfigured topology is a connected graph, treating the optical \dcn as a static network with occasional topology updates adapting to long-term traffic patterns~\cite{OSA, WaveCube, Flat-tree}. Google's Jupiter and Lightwave fabrics follow this approach, with reconfiguration intervals of minutes to hours~\cite{JupiterEvolving, Lightwave}.

\textbf{\tro optical \dcns}, in contrast, leverage advanced \ocs technologies for rapid circuit reconfiguration, optimizing latency for mice flows~\cite{RotorNet, Opera, Sirius, Shale, PULSE, RealizeRotorNet}.
Each \ocs configuration is held for a brief \textit{time slice}, typically lasting only microseconds or even sub-microseconds, with a fixed \textit{optical schedule} rotating through predefined topologies throughout time slices. These topologies maximize connectivity regardless of traffic patterns, often using expander graphs. 
The optical schedule repeats, each \textit{optical cycle} having tens of time slices to fully diversify network connectivity over time. %

\begin{figure}[tb]
    \centering%
    \includegraphics[width=\linewidth]
    {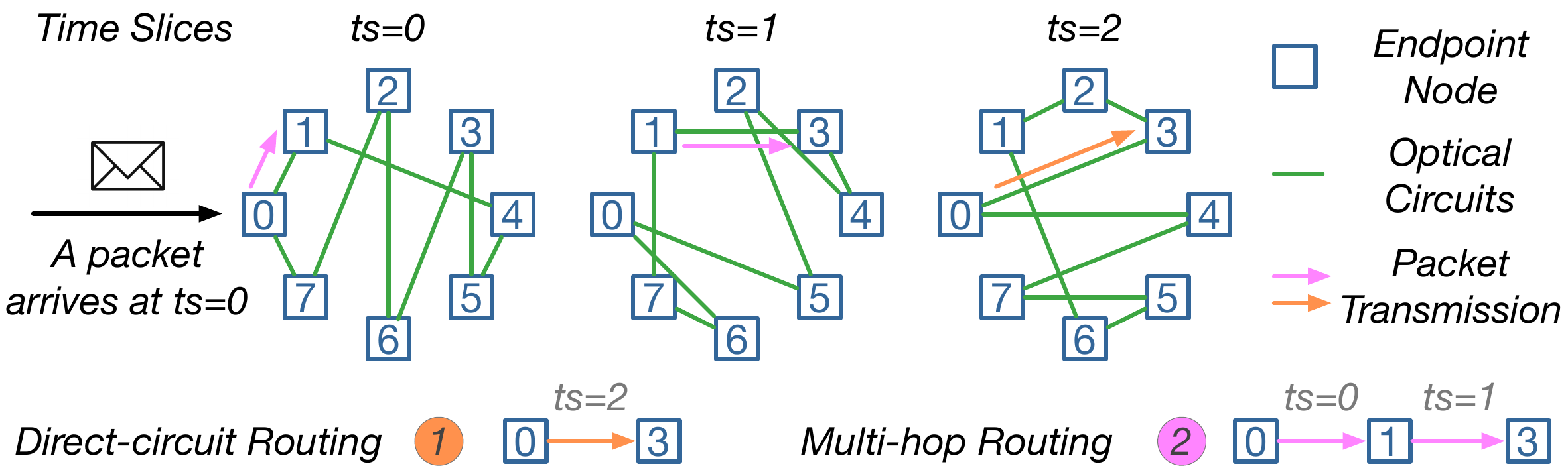}
    \figcap{A routing example in a \tro optical \dcn. The arrows represent the paths for a packet from endpoint node $N_1$ to $N_3$ arriving at time slice $ts$=0 under different routing schemes.}
    \label{fig:routing-example}
\end{figure}

\subsection{Routing in Optical \dcns}\label{sec:optical-routing}

Constrained to flow tables in the traditional SDN paradigm, routing in \tra architectures treats each reconfigured topology independently, with packets routed within the topology instance as if on a static topology.
In \textit{TA-1} architectures, mice flows are routed on the static topology, while elephant flows switch between the static topology and sporadic optical circuits.
The flow table on each electrical endpoint presents a default static route, and when a circuit becomes available, the optical controller updates it with a higher-priority route.
\textit{TA-2} architectures route traffic within each topology instance until it gets reconfigured, at which point the optical controller updates the flow table to shift traffic to the new paths.

Routing in \tro architectures is more challenging due to the ultrashort time slices,
where a packet in transit may cross multiple time slices, passing through different topologies.
This requires routing paths to be pre-computed offline, based on the fixed optical schedule, and electrical endpoints must execute the routing plan accurately for each time slice.

In Fig.~\ref{fig:routing-example}, for instance, a packet arriving at endpoint node $N_0$ at time slice $ts=0$ and destined to $N_3$ can either wait until $ts=2$ to take the \textit{direct} circuit \textcircled{1} when circuit $N_0 \leftrightarrow N_3$ appears, or can go through a more readily available \textit{multi-hop} path \textcircled{2} via the intermediate hop $N_1$.
Multi-hop paths can happen in the same time slice~\cite{Opera} or across time slices~\cite{RotorNet, Sirius, UCMP, HOHO}. In this example, path \textcircled{2} crosses two time slices: the packet reaches $N_1$ in $ts=0$ over circuit $N_0 \leftrightarrow N_1$, then waits until $ts=1$ for circuit $N_1 \leftrightarrow N_3$ to become available.

This time-based routing is fundamentally different from that of static networks, posing two additional requirements.

\textit{Req. 1}: To determine the time slice in which a packet arrives and map it to the corresponding routing path.

\textit{Req. 2}: To buffer the packet if it needs to be sent in a different time slice than the one in which it arrives.

As a general framework, \name must meet these new requirements of \tro architectures, while being compatible with the traditional SDN paradigm used by \tra architectures. This motivates us to propose the new time-flow table network abstraction, which, as shown in \S\ref{sec:api}, not only supports both \tra and \tro architectures but also extends \tra designs beyond the traditional SDN paradigm, enabling routing packets across different topologies.

\section{Time-Flow Table}\label{sec:table}

\begin{figure*}[t]
    \centering%
    \includegraphics[width=\linewidth]
    {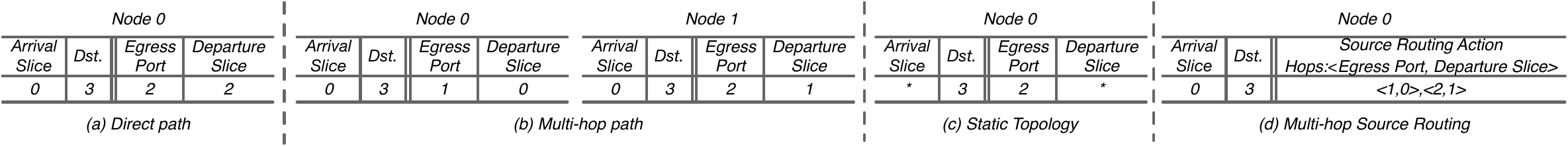}
    \figcap{Time-flow table examples for: \textbf{(a)} direct path \numcirc{1} and \textbf{(b)} multi-hop path \numcirc{2} in Fig.~\ref{fig:routing-example}, \textbf{(c)} a static path in an \tra architecture or traditional \dcn, and \textbf{(d)} source routing realization of path \numcirc{2}, i.e., equivalent of per-hop lookup in (b).}
    \label{fig:time-flow-table}
\end{figure*}

In this section, we introduce the key enabler of \name---the \textit{time-flow table} abstraction, with the system support for it described in \S\ref{sec:q-rotation}.
Compared to the traditional flow table, this abstraction highlights \textit{the notion of time} to meet the requirements of time-based routing presented above (\S\ref{sec:optical-routing}).

As shown in Fig.~\ref{fig:time-flow-table}, for \textit{Req.~1}, we define \textit{arrival time slice} as a \textit{match field}, allowing a switch or a host NIC to determine the time slice of a packet upon arrival and map it to the appropriate path; for \textit{Req.~2}, we define \textit{departure time slice} as an \textit{action field} to enable routing protocols that may dispatch packets in future time slices~\cite{RotorNet, HOHO, UCMP}. The time-flow table is backward-compatible with regular flow tables, if we set arrival and departure time slices to be wildcards.

Next, we use examples from Fig.~\ref{fig:time-flow-table} to demonstrate the expressiveness of the time-flow table abstraction in presenting diverse routing schemes of \tra and \tro optical \dcns, specifically with routing paths \textcircled{1} and \textcircled{2} from  Fig.~\ref{fig:routing-example}.

\para{Direct-circuit routing.}
Fig.~\ref{fig:time-flow-table}\,(a) shows the time-flow table of $N_0$ for the direct path \textcircled{1} from $N_0$ to $N_3$. A packet arriving at time slice $ts = 0$ must wait until $ts = 2$ for the direct circuit $N_0 \leftrightarrow N_3$, so the arrival time slice is set to 0, and the departure time slice is set to 2.

\para{Multi-hop routing.}
For the multi-hop path \textcircled{2} via $N_1$, the packet is forwarded immediately from $N_0$ to $N_1$ at $ts = 0$, so both the arrival and departure time slices in $N_0$'s time-flow table in Fig.~\ref{fig:time-flow-table}\,(b) are set to 0. At $N_1$, the packet arrives at $ts = 0$ (through circuit $N_0 \leftrightarrow N_1$) and waits until $ts = 1$ to be sent to $N_3$, so $N_1$'s time-flow table has an arrival time slice of 0 and a departure time slice of 1.

\noindent\textbf{Routing in \tra architectures and static \dcns.}
In Fig.~\ref{fig:time-flow-table}\,(c), the time-flow table reduces to a standard flow table when the arrival and departure time slices are set to wildcards, allowing packets to be forwarded immediately upon arrival. Routing in \tra architectures occurs in separate topology instances with static paths, similar to routing in a static irregular (non-Clos) topology, which can be presented with flow tables (refer~\S\ref{sec:optical-dcns}).
As the time-flow table reduces to a flow table, it supports \tra architectures as well as static \dcns.

\para{Source routing.}
We have shown examples of per-hop routing lookup so far.
Some routing schemes, however, cannot break down the paths into per-hop lookups and have to be implemented with source routing~\cite{Opera, UCMP}. Our time-flow table supports source routing by including the entire path in the action field at the source endpoint node, as a sequence of <egress port, departure time slice> tuples for each hop. For example, Fig.~\ref{fig:time-flow-table}\,(d) is the source routing equivalent of the per-hop tables in Fig.~\ref{fig:time-flow-table}\,(b), where the hops <1,0> and <2,1> for $N_0$ followed by $N_1$ are to be written to the packet.

\para{Multi-path routing.}
The time-flow table supports both per-packet and per-flow multi-path routing, as required by certain solutions~\cite{RotorNet, Sirius, Shale, UCMP}, through an optional path hashing field. For example, per-packet hashing can be based on timestamping or on the on-chip random number generator, while per-flow hashing using the five-tuple. 

The above examples cover existing routing primitives in \tra and \tro optical \dcns and traditional static \dcns. Future routing solutions will likely combine these primitives and can therefore also be supported by the time-flow table.

\section{Programming Model}\label{sec:program_model}

We introduced the time-flow table as a unified network abstraction to support both \tra and \tro optical \dcns with diverse routing schemes ($\S$\ref{sec:table}).
A user-friendly system should, however, relieve users of the tedious and error-prone task of managing low-level time-flow table entries and optical circuits.
We propose, hence, a unified programming model for \name to ease implementation of \tra and \tro architectures, and potential new ones that blur the \tra-\tro boundary.

\begin{figure}[tb]
    \centering%
    \includegraphics[width=\linewidth]
    {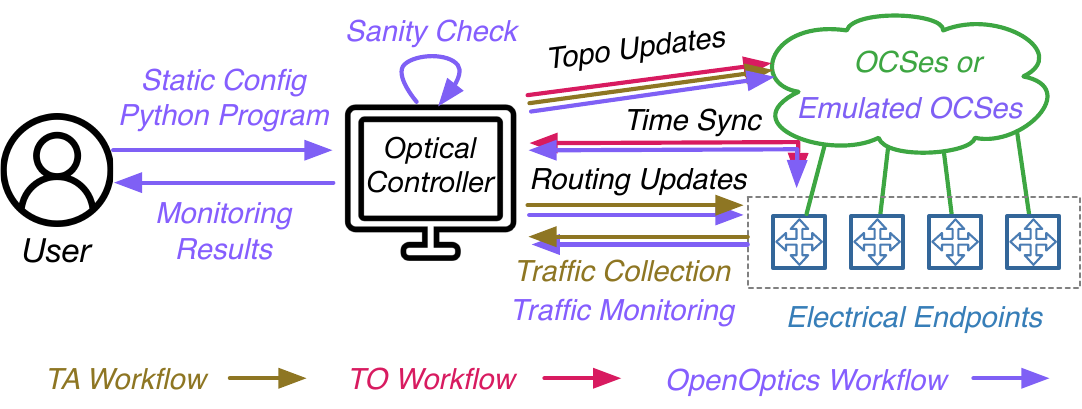}
    \figcap{\name system workflow.}
    \label{fig:workflow}
\end{figure}

\subsection{System Workflow}\label{sec:workflow}

A unified programming model requires a unified system workflow that combines the previously separate \tra and \tro workflows. The \name workflow is built on existing hardware components from \tra and \tro architectures, without introducing new components (Fig.~\ref{fig:workflow}). The only exception are the emulated \ocses on a programmable switch, which offer users access without the need for real \ocses. We also implement the entire workflow and system in Mininet (omitted from the figure) to make \name accessible to a broader range of users, such as students, without access to physical network hardware.

\para{\tra workflow.}
\tra architectures ($\S$\ref{sec:optical-dcns}) start with a default topology.
The electrical endpoints periodically report traffic statistics to the optical controller, using which the latter builds a global traffic matrix (TM).
The controller then optimizes the topology for that TM and computes routing paths for the optimized topology.
Next, it updates the flow tables (constrained by the legacy SDN paradigm, as established in $\S$\ref{sec:optical-routing}) on the electrical endpoints.
Once these routing changes are applied successfully, the controller reconfigures the topology by resetting the optical circuits.

\begin{table*}[ht]
\centering
\tabcap{\name user API functions, optional arguments in \textcolor{gray}{gray}.}\label{tab:api}
\footnotesize
\begin{tabular}{p{30pt}p{200pt}lll}
\hline
\textbf{Category}  & \textbf{APIs}  & \textbf{Description}    \\ \hline
\multirow{5}{*}{\textit{Topology}}
& \texttt{connect(Circuit<N1,port1,N2,port2,\textcolor{gray}{ts}>)$\rightarrow$bool} & Primitive function connecting \code{port1} \code{port2} of nodes \code{N1} \code{N2} in times slice \code{ts}. \\
& \texttt{topo(\textcolor{gray}{TM})$\rightarrow$[Circuit]} & Abstract function to generate \code{Circuit}s, with building block \code{connect()}. \\ 
& \quad\rotatebox[origin=c]{180}{$\Lsh$}\,\texttt{edmonds(TM),BvN(TM),jupiter(TM)
} & Circuit scheduling algorithms materializing \code{topo()} for \tra architectures. \\ 
& \quad\rotatebox[origin=c]{180}{$\Lsh$}\,\texttt{round\_robin(dimension,uplink)} & Optical schedule generation materializing \code{topo()} for \tro architectures. \\
& \texttt{deploy\_topo([Circuit])$\rightarrow$bool} & Check feasibility and deploy the topology. \\
\hline

\multirow{7}{*}{\textit{Routing}}
& \texttt{add(Entry<\textcolor{gray}{arr\_ts},src,dst,\textcolor{gray}{dep\_ts>},node)$\rightarrow$bool }
& Add time-flow table \code{Entry} on \code{node}; \code{arr\_ts}, \code{dep\_ts} to \code{null} turns a flow table. \\
& \texttt{routing([Circuit])$\rightarrow$[Path{<}src,dst,\textcolor{gray}{ts}{>}] }
& Abstract function to generate paths from \code{src} node to \code{dst} node at \code{ts}. \\
& \texttt{\quad\rotatebox[origin=c]{180}{$\Lsh$}\,direct(), ecmp(), wcmp(), ksp()}
& Routing algorithms materializing \code{routing()} for \tra architectures. \\
& \texttt{\quad\rotatebox[origin=c]{180}{$\Lsh$}\,vlb(), opera(), ucmp(), hoho()}
& Routing algorithms materializing \code{routing()} for \tro architectures. \\
& \texttt{neighbors(([Circuit],node,\textcolor{gray}{ts})$\rightarrow${[}node{]}}
& Helper function returning all nodes having direct circuits to \code{node} in \code{ts}. \\
& \texttt{earliest\_path([Circuit],src,dst,\textcolor{gray}{ts},max\_hop)$\rightarrow${[}Path{]} }
& Helper function returning paths from \code{src} to \code{dst} most recent to \code{ts} in \code{max\_hop}.\\
& \texttt{deploy\_routing([Path],LOOKUP,MULTIPATH)$\rightarrow$bool} & Compile \code{Path}s into \code{Entry}s and deploy, with \code{LOOKUP} and \code{MULTIPATH} options.\\
\hline

\multirow{3}{*}{\textit{Monitoring}}
& \texttt{collect(interval)$\rightarrow$TM} & Collect global traffic metric \code{TM} every \code{interval}. \\
& \texttt{buffer\_usage(node,port,interval)$\rightarrow$unsigned} & Query the buffer usage of \code{port} at \code{node} every \code{interval}. \\
& \texttt{bw\_usage(node,port,interval)$\rightarrow$unsigned} & Query the bandwidth usage of \code{port} at \code{node} every \code{interval}. \\
\hline
\end{tabular}
\end{table*}

\para{\tro workflow.}
With a pre-determined optical schedule, \tro architectures ($\S$\ref{sec:optical-dcns}) require minimal interaction between electrical endpoints and the optical controller. The controller reconfigures the \ocses per time slice to realize different topologies in the optical schedule. Routing paths for each topology are pre-computed offline and, according to the \tro proposals, pre-loaded into the electrical endpoints, which synchronize with the optical controller to self-regulate the time slices and enforce the appropriate routing paths. Without time-flow table abstraction and the supporting queue management system, however, how to enforce routing changes per time slice remains largely unknown; it is one of the reasons why most \tro proposals are not yet implemented.\footnote{Without time-based queue management, RotorNet~\cite{RealizeRotorNet}, the only implemented \tro architecture, is constrained to continuous routes.
Its routing lookup per topology extends flow tables with an additional match field.
}

\name supports \tro architectures with the time-flow table ($\S$\ref{sec:table}) and the corresponding queue management system ($\S$\ref{sec:q-rotation}) for time-based packet scheduling. As detailed in $\S$\ref{sec:q-rotation}, time-flow tables reside on electrical endpoints, such as switches and NICs with packet processing programmability, and the queue management system handles packet lookup, scheduling, buffering, and dispatching on a per-time-slice basis. Time synchronization solutions of existing \tro proposals either rely on specific optical hardware~\cite{Sirius} or achieve poor accuracy sufficient for particular architectures only~\cite{RealizeRotorNet}. We built a general hardware-independent synchronization protocol with nanosecond precision to support diverse architectures, which is described in a separate paper~\cite{OpSync}.

\name supports both traffic-based real-time topology updates (as in \tra) and batch topology configurations (as in \tro).
Breaking the \tra-\tro boundary enables novel hybrid designs (\S\ref{sec:api-examples}), such as a periodically updated optical schedule that reflects traffic evolution, or \tra and \tro subnetworks in different hierarchies working in tandem. 

\name enhances user interaction through API functions ($\S$\ref{sec:api}). Users specify high-level network behavior via a static configuration (\code{json} file) for hardware setups (e.g., \ocses count and structure, optical uplinks per endpoint, and time slice duration), along with a Python program that invokes the API functions. The optical controller sanity checks the inputs, e.g., test the feasibility of circuits and routing paths, and compiles them into circuit connections and time-flow table entries. We extend the basic traffic collection functionality in the \tra workflow to rich traffic monitoring APIs, providing network telemetry beyond traffic volume, e.g., buffer and link usage, to monitor network health.

\subsection{User API}\label{sec:api}

A user first creates an \name network object and then calls the topology, routing, and monitoring APIs listed in Tab.~\ref{tab:api}.
The API calls accept optional arguments for each architecture type to offer a unified workflow. For instance, \tra architectures set the time slice \code{ts=null} to operate within each topology configuration, and \tro architectures set the traffic matrix \code{TM=null} to disregard traffic.

\para{Topology APIs.}
We define the primitive call \code{connect()} to establish a circuit between endpoint nodes.

The abstract function \code{topo()} builds on \code{connect()} to realize diverse topologies. We materialize it to implement \tra circuit scheduling algorithms, such as the classic Edmonds' matching and Birkhoff-von-Neumann algorithms used in c-Through~\cite{cThrough} and Mordia~\cite{Mordia}, and the gradual evolving approach used in Jupiter~\cite{JupiterEvolving}. We also implement round-robin variants for
\tro optical schedules. For instance, Shale~\cite{Shale} uses a three-dimensional round-robin with a single optical uplink per node, and Opera~\cite{Opera} uses a single-dimensional round-robin with $N$ uplinks per node.
Users can define custom topologies by overriding \code{topo()}.

The action function \code{deploy\_topo()} compiles the node-level circuits into \ocs internal connections based on the \ocs structure specified in the static configuration file. The optical controller verifies the feasibility of the physical circuits and deploys them onto the \ocses.

\begin{figure*}[t]
\centering
\begin{minipage}[htbp]{0.22\linewidth}
\begin{minted}[fontsize=\scriptsize,linenos,xleftmargin=8pt]{python}
#config={"node":"host",
#  “node_num”:128,
#  "uplink":2,
#  "IPs":["10.0.0.1",..]}
net=OpenOptics.net(config)
circuits=round_robin(
  dimension=1,
  uplink=config.uplink)
paths=vlb(circuits)
net.deploy_topo(circuits)
net.deploy_routing(paths,
  LOOKUP="hop",
  MULTIPATH="packet")
\end{minted}
\sfigcap{ RotorNet (\tro).}\label{fig:code_rotornet}
\end{minipage}
\begin{minipage}[htbp]{0.25\linewidth}
\begin{minted}[fontsize=\scriptsize,linenos,firstnumber=1,xleftmargin=10pt]{python}
net=OpenOptics.net(config)
circuits=jupiter(
  TM=null)
paths=wcmp(circuits)
net.deploy_topo(circuits)
net.deploy_routing(paths)
while(
  TM=net.collect("24h")):
  circuits=jupiter(
    TM, circuits)
  paths=wcmp(circuits)
  net.deploy_routing(paths)
  net.deploy_topo(circuits)
\end{minted}
\sfigcap{ Jupiter (\tra).}\label{fig:code_jupiter}
\end{minipage}
\begin{minipage}[htbp]{0.23\linewidth}
\begin{minted}[fontsize=\scriptsize,linenos,firstnumber=1,xleftmargin=10pt]{python}
net=OpenOptics.net(config)
circuits=round_robin(
  1,config.uplink)
paths = vlb(circuits)
net.deploy_topo(circuits)
net.deploy_routing(paths)
while(
  TM=net.collect("10min")):
  # Custom optical schedule
  circuits=sorn(TM)
  paths=vlb(circuits)
  net.deploy_routing(paths)
  net.deploy_topo(circuits)
\end{minted}
\sfigcap{Semi-oblivious ({\tra}+{\tro}).}\label{fig:code_semi}
\end{minipage}
\begin{minipage}[htbp]{0.25\linewidth}
\begin{minted}[fontsize=\scriptsize,linenos,firstnumber=1,xleftmargin=10pt]{python}
#config={“node”:"rack",
# "node_num":128,...};rack_conf={
# “node”:"host","node_num":64,...}
net=OpenOptics.net(config)
for rack in net.nodes:
  r_cts=round_robin(
    1,rack_conf.uplink)
  rack.deploy_topo(r_cts)
  rack.deploy_routing(vlb(r_cts))
while(TM=net.collect("1h")):
  cts=BvN(TM)
  net.deploy_topo(cts)
  net.deploy_routing(wcmp(cts))
\end{minted}
\sfigcap{Hierarchical ({\tra}+{\tro}).}\label{fig:code_hierarchy}
\end{minipage}
\figcap{Code snippets of implementing various optical architectures using the \name APIs (\S\ref{sec:api}).}\label{fig:snippets}
\end{figure*}

\para{Routing APIs.}
The basic function \code{add()} allows users to add time-flow table entries directly, e.g., for debugging purposes. It supports flow table entries where the arrival and departure time slices are both set to \code{null} as wildcards (\S\ref{sec:table}).

We define an abstract function \code{routing()} to implement diverse routing algorithms, which return the set of paths per source-destination node pair, and additionally per time slice for \tro architectures.
We materialize it for \tra routing algorithms running within each topology configuration\, including direct-circuit routing~\cite{RotorNet}, ECMP~\cite{Fat-tree}, WCMP~\cite{JupiterEvolving}, k-shortest path~\cite{Flat-tree}, and \tro algorithms across topologies, including 
VLB~\cite{Sirius, RotorNet}, Opera~\cite{Opera}, UCMP~\cite{UCMP}, and HOHO~\cite{HOHO}.

Users can implement customized routing algorithms via overriding functions, and we provide helpers to simplify them. \code{neighbors()} retrieves, for instance, connected neighbors for a node in a time slice, useful for VLB, and \tra routing algorithms on a topology instance when \code{ts=null}; similarly, \code{earliest\_path()} finds the first path between a source and destination node since a given time slice, facilitating direct-circuit routing, UCMP, and HOHO, and shortest path routing is a special case of it with \code{ts=null} in one topology. 

Finally, \code{deploy\_routing()} compiles the paths into time-flow table entries and loads them onto each node. It supports \code{LOOKUP} types of per-hop and source routing by decomposing the path into next-hop nodes or retaining the entire path in the action field. It also supports packet- and flow-level \code{MULTIPATH} by hashing the ingress timestamp or five tuples.

\para{Monitoring APIs.}
We offer telemetry functions, such as \code{buffer\_usage()} and \code{bw\_usage()} for querying the buffer and bandwidth usage of a port on a node. The \code{collect()} function for traffic collection in \tra architectures is considered a special case of monitoring. New monitoring functions can be easily added as needed.

\subsection{Example Programs}\label{sec:api-examples}

Fig.~\ref{fig:snippets} shows how the unified workflow (\S\ref{sec:workflow}) and APIs (\S\ref{sec:api}) simplify implementation of \tra and \tro architectures, and novel hybrid designs with a few lines of Python code. 

\para{RotorNet (\tro).}
An \name network object is first created with a static configuration, using which the optical controller connects to the nodes, in this case hosts.
RotorNet~\cite{RotorNet}, being a \tro architecture, uses a single-dimensional round-robin optical schedule with the number of optical uplinks per node specified in the configuration file, and it runs VLB routing on the optical schedule. The generated topologies and paths are deployed onto the \ocses and host NICs, where time-flow table entries are created for per-hop lookups, as opposed to source routing, enabling packet-level multipath routing for packet spraying in VLB.

\para{Jupiter (\tra).}
The \tra architecture Jupiter~\cite{JupiterEvolving} starts with a uniform mesh topology, an empty TM, and WCMP routing.
Once every 24 hours, a new TM is collected and the topology optimized. The computed routing changes are deployed as higher-priority routes atop existing ones before reconfiguring the topology, ensuring seamless traffic switches.

\para{Semi-oblivious ({\tra}+{\tro}).}
\name enables new architectural designs beyond the traditional \tra-\tro boundary, thus naturally supports the recent semi-oblivious proposal that builds skewed round-robin optical schedules to reflect traffic patterns~\cite{Semi-oblivious}.
It starts with a single-dimensional round-robin optical schedule (Fig.~\ref{fig:code_semi}) and VLB routing as a regular \tro network, and the topology and routing are updated every 10 minutes based on the new TM. Our topology API \code{topo()} can be easily extended to support the custom topology-building algorithm \code{sorn()}, which modifies \code{round\_robin()} by creating dense connections between hotspot nodes and sparse ones elsewhere. This design is traffic-driven like \tra architectures, but each reconfiguration loads an optical schedule with a batch of topologies like \tro architectures.

\para{Hierarchical ({\tra}+{\tro}).}
We can envision other types of \tra-\tro hybrid designs, such as the hierarchical network in Fig.~\ref{fig:code_hierarchy}. For emerging ML workloads, GPU machines within a rack can be interconnected through a \tro scale-up network, leveraging its reach connectivity, while ToRs can be further interconnected through a \tra scale-out network to manage traffic locality across racks. We define separate static configuration files for the two network levels (Fig.~\ref{fig:code_hierarchy}), and a network object is first created for the core inter-rack network. Then, for each node in the core network, an intra-rack \tro network is created, similar to RotorNet (Fig.~\ref{fig:code_rotornet}). The inter-rack network then collects traffic and updates topology and routing like Jupiter (Fig.~\ref{fig:code_jupiter}).

\section{\name System}\label{sec:tor-sys}

In this section, we introduce the backend system for the time-flow table ($\S$\ref{sec:table}) and programming model ($\S$\ref{sec:program_model}), which are general to both switch-centric and host-centric optical \dcns with switches and hosts connected to the \ocses, respectively. Our description is based on our current implementation for the mainstream switch-centric design for scalability, using Intel Tofino2 programmable switches and \vma userspace library~\cite{VMA} on Mellanox NICs. 
Our implementation applies to both ToR-based and pod-based switch-centric designs, where our switch system is deployed on ToRs or pod switches, and host system on end hosts. Other switches in a pod need no change.
The host-centric version of \name with Corundum FPGA-based NICs~\cite{Corundum} is under development.

The main system components include the customized queue management system to support the time-flow table ($\S$\ref{sec:q-rotation}) and infra services that provide common features and optimization knobs for various optical architectures ($\S$\ref{sec:infra_services}). We also outline system peripherals, including the emulated optical fabric and Mininet education toolkit
($\S$\ref{sec:mininet}).

\begin{figure}[tb]
    \centering
    \includegraphics[width=0.99\columnwidth]
    {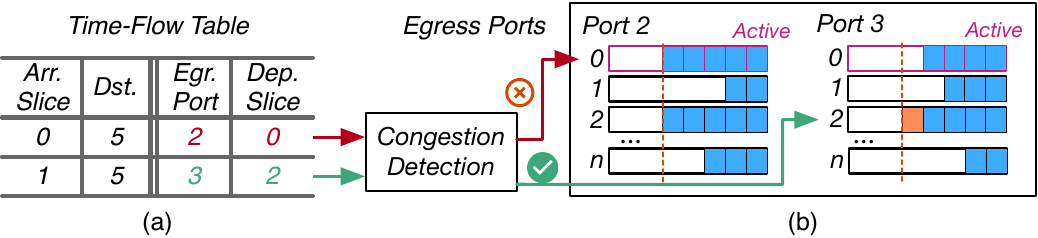}
    \figcap{Example of packet processing on a switch, assuming the active calendar queue is \code{q=0} for time slice \code{ts=0}.
    }\label{fig:tor-example}
\end{figure}

\subsection{Time-Based Queue Management}\label{sec:q-rotation}

The time-flow table requires time-based packet scheduling to dispatch packets at the planned departure time slice, and buffering is needed if the departure time slice is later than the arrival time slice, both go beyond  priority-based packet scheduling in traditional queue management systems. 
We leverage the \textit{queue pausing} feature of modern programmable switches~\cite{lee2020tofino2} to rearchitect the queue management system, which allows for controlled pausing and resuming of queues.
If egress queues are assigned to specific time slices, packets can be buffered in queues corresponding to their departure time slices, which are paused until the expected time to release the packets.

We implement this design inspired by the calendar queues framework~\cite{CalendarQueue}.
A calendar queue is associated with a ``calendar day'', and packets can be enqueued at a ``rank'' for a future day.
A \textit{calendar day} is \textit{a time slice} in our case.
We create a set of calendar queues per egress port, assigning each time slice a queue sequentially until queue exhaustion to wrap around. The \textit{rank} of an ingress packet is the difference between its \textit{departure time slice} and \textit{arrival time slice}.

Queue pausing and resuming can be triggered by any ingress packet in the data plane. We control this process using the on-chip packet generator available in programmable switches~\cite{joshi2019timertasks}, which reliably sends a packet into the ingress pipeline at a specified time interval. We configure this interval to match the \textit{time slice duration}. Recall \name synchronizes electrical endpoints, switches and host NICs included, with the optical controller at nanosecond precision~\cite{OpSync}, so the on-switch packet generator can send a packet at the start of each time slice to initiate \textit{queue rotation} across all egress ports. This rotation pauses the currently active queue and resumes the queue for the upcoming time slice. Each switch keeps track of the \textit{active queue} for the current time slice, which is consistent across all egress ports.

Fig.~\ref{fig:tor-example} illustrates two sets of calendar queues at egress ports \code{p=2} and \code{p=3}. At time slice \code{ts=0}, the active queue is \code{q=0} for all ports.
An incoming packet matching the first entry in Fig.~\ref{fig:tor-example}\,(a) is mapped to \code{q=0} of \code{p=2} in Fig.~\ref{fig:tor-example}\,(b), because its departure time slice equals the arrival time slice, meaning it should be enqueued in the active queue for immediate transmission. 
When the time slice advances to \code{ts=1}, queue rotation occurs, making \code{q=1} the active queue. Another packet arriving at this time matches the second entry and should be enqueued in \code{q=2} of \code{p=3}, to be sent out one time slice later.

Queue pausing is supported on some FPGA-based NICs, making our queue management system easily portable to host-centric optical \dcns. For \tra architectures and static \dcns that rely on traditional flow tables, we can simply disable calendar queues and revert to the default queue settings.

\subsection{Infrastructure Services}\label{sec:infra_services}

We abstract common functionalities and optimization options from existing optical architectures as infra services, allowing users to focus on exploring upper-layer network protocols instead of reinventing the wheel. Most of these features are proposed but not implemented, with a few implemented in ways specific to particular architectures. Our infra services enable complete and straightforward implementation of these proposals, and future ones.

\para{Congestion detection.}
Many optical architectures adopt congestion control (CC) mechanisms~\cite{Opera, UCMP, HOHO, TDTCP, reTCP}, yet congestion detection in calendar queues ($\S$\ref{sec:q-rotation}) is non-trivial.
An optical circuit transmits a fixed amount of data per time slice. During congestion, packets may miss their scheduled time slice and remain paused in the calendar queue for a full cycle, leading to excessive delays. This occurs when a calendar queue accumulates more data than it can transmit within a time slice. Under short time slices, this threshold of accumulated data---determined by the remaining time in the time slice---may fall below the congestion threshold typically used in CC protocols for regular queues, introducing an additional condition for congestion detection.

It is challenging to determine if an incoming packet can fit into the intended calendar queue, as commercial switches cannot access the occupancy of calendar queues (egress queues) in the ingress pipeline before enqueuing.\footnote{Tofino2's ``ghost thread'' feature claims to support this functionality~\cite{lee2020tofino2}, but our tests show queue occupancy readings are outdated by milliseconds, insufficient for microsecond-scale time slices in \tro architectures.}
We thus implement a \textit{queue occupancy estimation} method using a register array in the ingress pipeline to track each calendar queue's occupancy. Since registers can only be updated by ingress packets, we increase the occupancy when packets are enqueued and use a packet generator to create an ingress packet every \textit{update interval} to reduce the occupancy, assuming line-rate packet dequeuing. Evaluation in Fig.~\ref{fig:es_error} shows $\uns{50}$ update interval results in less than one packet estimation error with minimal switch pipeline processing overhead. More details of this design is explained in Appx.~\ref{sec:slice-miss}.

A calendar queue is full if its occupancy exceeds the admissible data amount for the elapsed time of the time slice (bandwidth $\times$ time). Congestion occurs if (1) the calendar queue is full or (2) the congestion threshold is reached, whichever happens first. This service detects congestion while giving users flexibility in how to respond. For example, in Fig.\ref{fig:tor-example}, an incoming packet in \code{ts=0} that matches the first table entry finds the intended calendar queue \code{q=0} on port \code{p=2} full. The architecture can then trigger its own CC mechanism, such as packet dropping, packet trimming~\cite{Opera}, or deferring the packet to a later time slice~\cite{HOHO, UCMP}. In contrast, a packet arriving at \code{ts=1} that matches the second entry can be enqueued safely in \code{q=2} of \code{p=3}, which is not full yet.

\para{Traffic push-back.}
While we give users flexibility in choosing CC solutions, we offer optional traffic push-back as a last-resort protection if CC fails.
When a packet detects its designated calendar queue is full, it and all subsequent packets to that queue should be rejected.
If the push-back service is enabled, a rejected packet triggers a \textit{push-back} message containing the time slice of the queue. This message is broadcast 
to the sender ToR or pod switch, which forwards it to all connected hosts, preventing them from sending traffic to the destination switch during that time slice.
This broadcast increases push-back coverage to hosts that might attempt to send to the same queue.
Traffic is blocked on the hosts using the flow pausing service, which will be introduced next.

\para{Flow pausing.}
Many \tra architectures pause elephant flows at the source until they can be sent over direct circuits~\cite{Helios, cThrough, OSA, Mordia, ProjectToR, WaveCube}. In \tro architectures, while calendar queues (\S\ref{sec:q-rotation}) provide buffering, elephant flows still place a significant burden on switch buffers. It is desirable to also route elephant flows over direct circuits as an optimization. We implement flow pausing on hosts to support this.

We utilize \textit{flow aging}~\cite{PIAS} to identify elephant flows without explicit flow size information.
ToRs or pod switches broadcast \textit{signal messages} to the connected hosts, notifying them of upcoming circuit connections.
We implement flow pausing with the user-space \vma library to achieve high performance and transparency to TCP/UDP applications~\cite{VMA}.
\vma links sockets to the userspace \code{lwIP} stack,
where we intercept socket send calls to suspend sending out data from the segment queue until the circuit is ready.
Suspending and resuming applications require no additional memory buffers beyond the segment queue, as applications are naturally pushed back by the socket interface when the segment queue reaches its capacity. Fig.~\ref{fig:transport} demonstrates high throughput of our implementation.

\para{Traffic collection.}
A key feature of \tra architectures is traffic collection.
Architectures that reconfigure topology at seconds to minutes frequency establish circuits to serve elephant flows~\cite{Helios, cThrough, ProjectToR, WaveCube, OSA}, so we leverage flow pausing on hosts, with packets buffered in separate queues inside \code{vma} based on the destination switch. Hosts periodically report traffic volume per destination switch to their connected switches, which then aggregate and relay the data to the optical controller.
For infrequent topology reconfigurations on an hourly scale, as in Jupiter~\cite{JupiterEvolving}, the network operates on a static topology most of the time. Switches simply track the total traffic sent to each destination switch and report the statistics to the optical controller.

\para{Buffer offloading.}
Some multi-hop routing schemes consume excessive buffer space at intermediate switches, such as VLB where packets may wait an entire optical cycle before being forwarded. Some architectures suggest offloading packet buffering from switches to the connected hosts~\cite{RotorNet, Opera}. Although this proposal has not progressed beyond simulation, \name, as a general framework, implements it to enable full realization of these architectures and support more demanding buffering needs for future designs.

Each switch only keeps $N$ calendar queues per egress port for the immediate future and stores the rest for later time slices onto hosts under it.
As the time slices corresponding to the host-resident calendar queues approach, the packets are sent back to the switch in advance, guided by circuit notification messages. To keep the logic on switches lightweight, they randomly select hosts to balance the load and delegate bookkeeping to hosts, which initiate returning of offloaded packets.
We implement buffer offloading also with \vma, dedicating an application per host to isolate it from the main data path and ensuring low latency.
The offloaded packets are stored similar to paused flows as described above.

\begin{figure}[tb]
    \centering
    \begin{subfigure}[t]{0.65\columnwidth}
        \includegraphics[width=\linewidth]{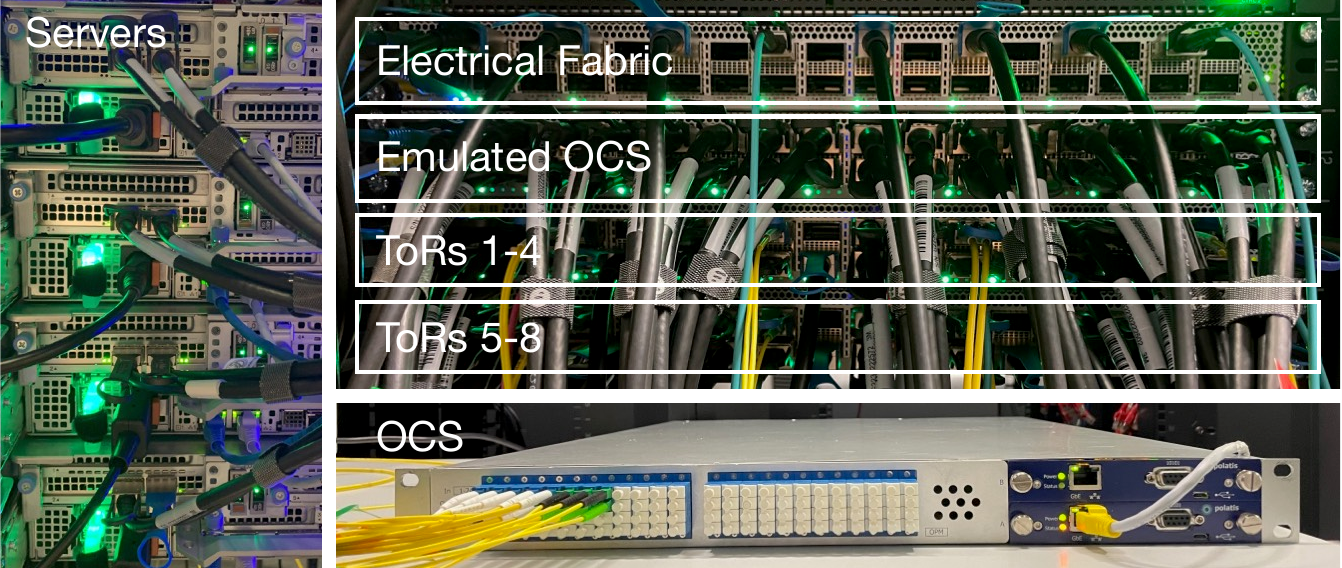}
        \figcap{Testbed photo.}
        \label{fig:testbed_pic}
    \end{subfigure}
    \hfill
    \begin{subfigure}[t]{0.34\columnwidth}
        \includegraphics[width=\linewidth]{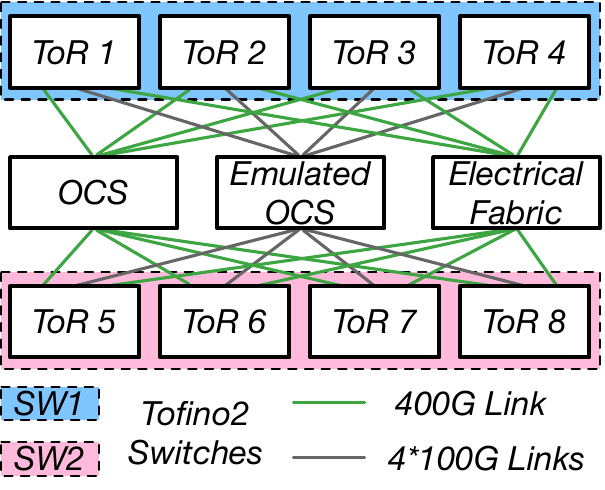}
        \figcap{Testbed diagram.}
        \label{fig:testbed_diagram}
    \end{subfigure}
    \figcap{Testbed setup (servers in~\ref{fig:testbed_pic} are omitted from~\ref{fig:testbed_diagram}).}\label{fig:testbed}
\end{figure}

\subsection{Emulated Optical Fabric and Mininet}\label{sec:mininet}

\para{Emulated optical fabric.}
To extend \name to users without real \ocses, we provide an emulated optical fabric using P4 programmable devices. It abstracts diverse \ocs structures of \ocses (Fig.~\ref{fig:optical-dcn-example}) as a single logical \ocs, maximizing connectivity with limited device ports. This logical \ocs offers time-based connectivity between electrical endpoints, with configurable circuit reconfiguration delays and circuit durations. To enhance realism, we enable the ``cut-through'' mode on commercial devices to minimize processing delays, closely approximating physical \ocs performance.

The emulated optical fabric is synchronized with the optical controller, and
circuit on-offs are emulated using a lookup table. Available circuits are realized by direct-through entries between the connected ports, while packets over disconnected circuits fail to match any entries and are dropped. 
The topology can be configured through the topology APIs in Table~\ref{tab:api}, then the optical controller programs the lookup table to update the topology for \tra architectures, and for \tro architectures loads the entire optical schedule, with time slices corresponding to specific circuits.
The circuit reconfiguration period is managed with a special (usually short) time slice, during which all affected packets are dropped.

\para{Mininet toolkit.}
We offer a fully emulated version of \name on Mininet for users without physical network equipment. While \name' time-flow table and system implementation are portable to various P4-capable environments, its novel time-based packet scheduling (\S\ref{sec:q-rotation}) is not natively supported by the Mininet switch stack. Thus, we implement queue pausing on the \code{BMv2} open-source P4 software switch~\cite{bmv2}, by enabling processes to pull packets from egress queues only during specified time periods, and migrate the \name queue management system.
This emulated version includes all \name components in Fig.~\ref{fig:workflow}, with the optical controller as a Mininet application, and the electrical endpoints and emulated optical fabric running on \code{BMv2}. Users interact with the optical controller using the same static configurations and APIs as in the physical system.

\section{Research Use Cases}\label{sec:research}

Below, we discuss \name{}'s ability to enable novel, previously unachievable, research use cases.

\para{Testbed.}
To demonstrate the capabilities of \name, we build a testbed with both real and emulated optical fabrics, as well as a traditional electrical fabric.
As shown in Fig.~\ref{fig:testbed}, the setup includes a Polatis Series 6000 MEMS \ocs, four EdgeCore DCS810 Intel Tofino2 switches, and four servers equipped with Mellanox ConnectX-5 \uGbps{100} dual-port NICs.
The MEMS \ocs functions as an optical network fabric.
Two Tofino2 switches are each divided into four logical ToRs (Fig.~\ref{fig:testbed_diagram}), while the third Tofino2 switch emulates another optical fabric
with flexible structures of emulated \ocses, and the fourth serves as an electrical fabric.
Each of the eight ToRs connects to the OCS and the electrical fabric with a \uGbps{400} link and to the emulated optical fabric with four \uGbps{100} links to support flexible emulation. 
The servers' dual-port NICs provide eight \uGbps{100} links, each connected to a ToR to work as eight individual hosts.

\para{Traffic.}
We run both latency-sensitive and throughput-intensive applications on the testbed to examine mice and elephant flows. We use the \textit{Memcached}~\cite{Memcached} key-value storage for the latency-sensitive application, running one Memcached server and seven Memslap\cite{Memslap} benchmarking clients each on a host. The clients  perform SET operations at milliseconds intervals, writing \uKB{4.2} of data to the server. As for the throughput-intensive application, we run \textit{ring allreduce} on the hosts using the \textit{Gloo} collective communication library~\cite{Gloo},
with varying data sizes from \uKB{800} to \uMB{20}.

\para{Case I: realistic comparison of architectures.}
We implement representative optical architectures on \name to validate system correctness and demonstrate side-by-side performance comparisons enabled by \name.
Particularly, we implemented \textit{c-Through}~\cite{cThrough}, \textit{Jupiter}~\cite{JupiterEvolving}, and \textit{Mordia}~\cite{Mordia} from the \tra class, and \textit{RotorNet}~\cite{RotorNet} and \textit{Opera}~\cite{Opera} from the \tro class. We also realized the traditional \textit{Clos}~\cite{Fat-tree} network for baseline comparison.

Clos relies on the electrical network fabric, while the MEMS \ocs, with tens of milliseconds long reconfiguration delays, supports Jupiter and c-Through. As a hybrid architecture, c-Through also connects to the electrical fabric, rate-limited to \uGbps{10} for consistency with the original design. Mordia, RotorNet, and Opera, requiring finer-grained circuit reconfiguration, use the emulated optical fabric. We apply the native routing schemes and optical fabric settings for these architectures and also run \textit{UCMP}~\cite{UCMP} routing on top of RotorNet to show the latest performance of \tro architectures.

\begin{figure}[t]
    \centering
    \includegraphics[width=0.9\linewidth]{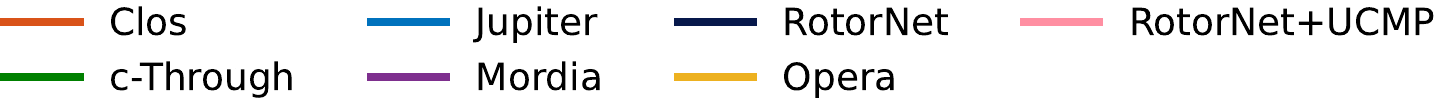} %

    \begin{subfigure}[t]{0.49\columnwidth} \centering 
        \includegraphics[height=3.05cm]{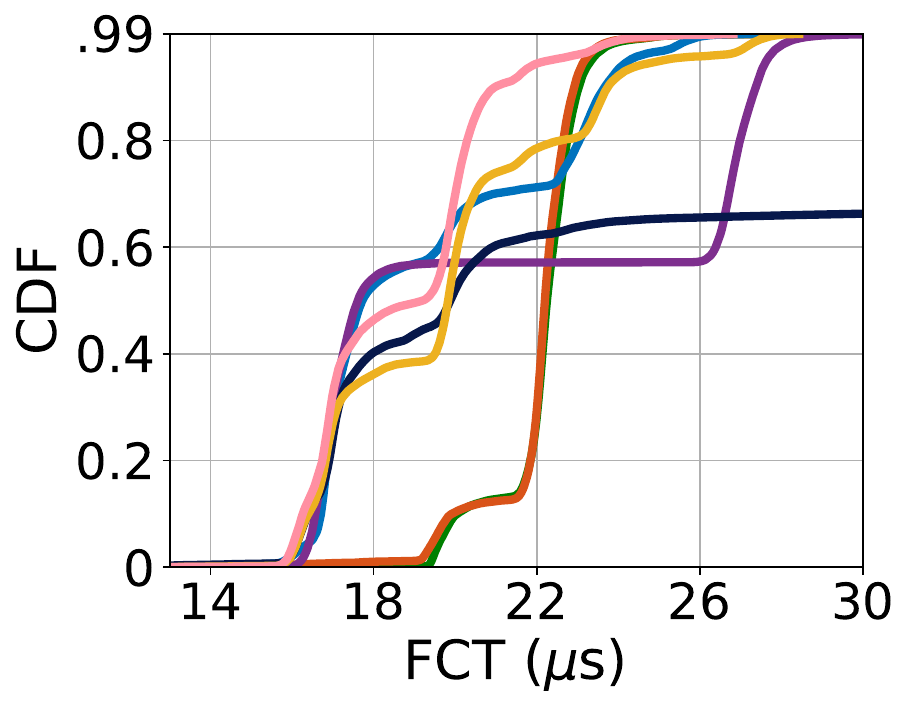}
        \phantomcaption\label{fig:memcached}
    \end{subfigure}
    \hfil
    \begin{subfigure}[t]{0.49\columnwidth} \centering 
        \includegraphics[height=3.05cm]{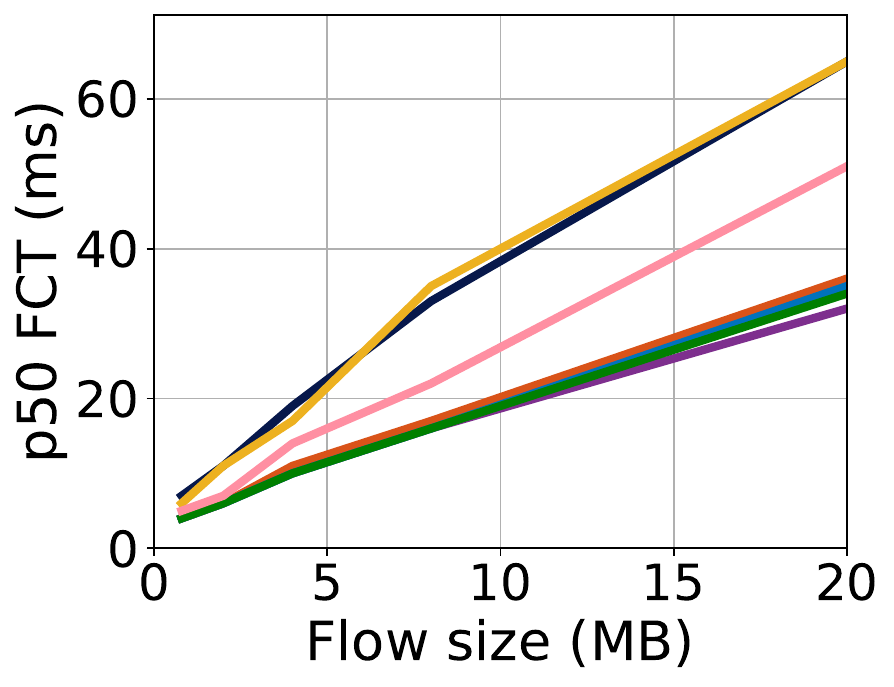}
        \phantomcaption\label{fig:allreduce_longflow}
    \end{subfigure}

    \figcap{FCTs of
    (a) Memcached and (b) Gloo allreduce.}\label{fig:applications}
\end{figure}

The distribution of flow completion times (\fcts) we obtain using \name for mice and elephant flows (Fig.~\ref{fig:applications}) match the \fct trends reported in the concerned prior work (that are either proprietary or only partially realized), which confirms the correctness of our implementation.
We augment this validation later (\S\ref{sec:perf}) by reproducing published results.

In Fig.~\ref{fig:memcached}, c-Through shows similar mice flow \fcts to Clos. As a hybrid architecture, it uses the optical fabric only for elephant flows, while sending mice flows over electrical network, which has minimal impact on their \fcts. Jupiter improves \fcts by offering an optimized topology with fewer hops tailored to the traffic. Mordia, also a \tra design, establishes circuits on demand.
This results in low \fcts for flows with immediately available circuits, but a long tail otherwise.

\tro architectures such as RotorNet and Opera are more sensitive to routing.
RotorNet employs VLB, which introduces significant circuit-waiting delays by waiting at intermediate hops, resulting in long tail \fct. Opera, in contrast, has low \fcts utilizing longer but always-available paths. UCMP lowers \fcts further by reducing path length.

For elephant flows in Fig.~\ref{fig:allreduce_longflow}, \tra architectures c-Through, Jupiter, and Mordia exhibit similar \fcts as Clos. They establish a ring topology using optical circuits that matches the traffic perfectly. 
\tro architectures RotorNet and Opera, however, double the \fcts as the circuits are available only half the time. UCMP improves throughput with more efficient routes, leading to reduced \fcts.

\para{Case II: investigation of the transport layer.}
Transport performance in optical \dcns is a new research focus due to challenges with dynamic paths, such as reTCP~\cite{reTCP} and TDTCP~\cite{TDTCP} that address bandwidth disparity in hybrid electrical-optical networks. Both were evaluated on the Etalon emulator~\cite{reTCP} for hybrid architectures only, limiting research in other architectural settings. This case study demonstrates transport performance analysis on \tro architectures using \name' multi-architecture support.

We measure the throughput of long-lasting \code{iperf3} flows between hosts on Clos, RotorNet with VLB and direct-circuit routing, and finally a hybrid version of RotorNet with \uGbps{100} bandwidth through the optical fabric and \uGbps{10} bandwidth through the electrical fabric, as in TDTCP.

\begin{figure}[tbp]
    \centering
    \includegraphics[width=\linewidth]{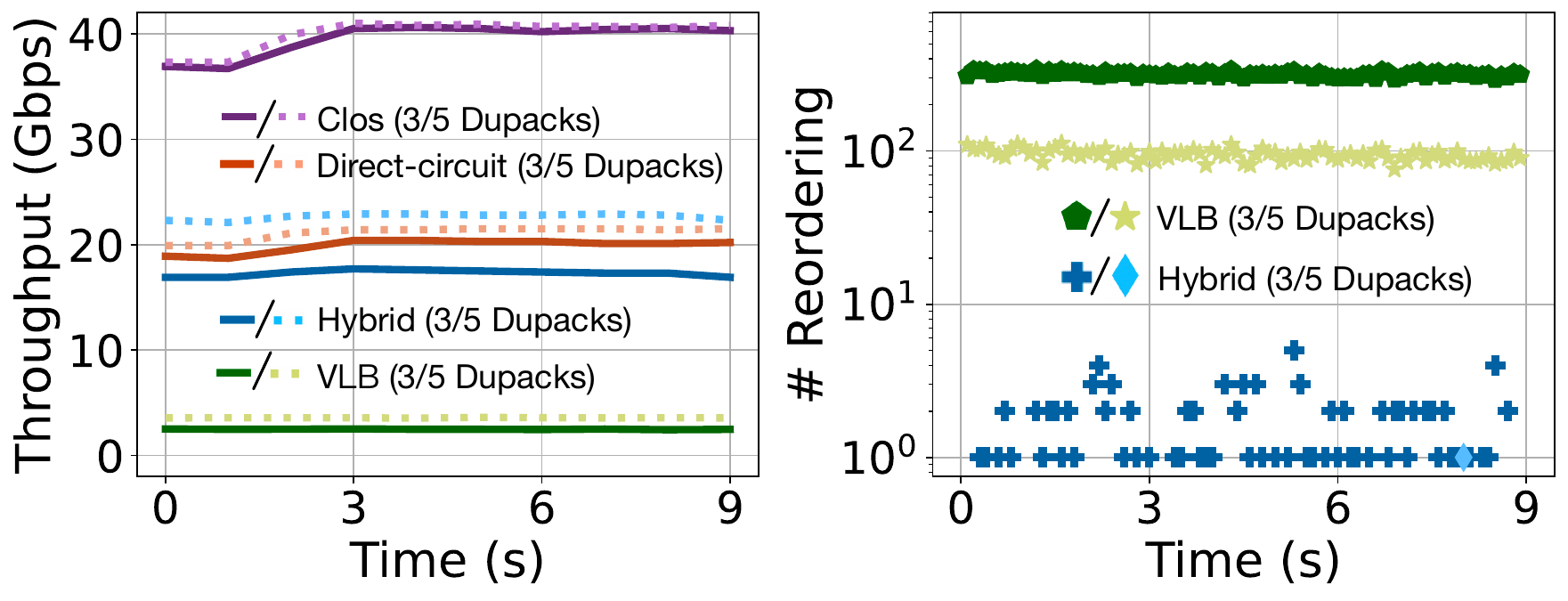} %
    \figcap{(a) TCP throughput and (b) number of packet reordering events with iperf traffic.
    }\label{fig:transport}
\end{figure}

In Fig.~\ref{fig:transport}(a), the \uGbps{40} throughput in Clos is the upper bound for \code{iperf3} on our testbed because it is CPU-bound. As expected, for direct-circuit routing, our host system with flow pausing (\S\ref{sec:infra_services}) achieves roughly half that throughput due to the circuit being available 50\% of the times.
In contrast, VLB exhibits significantly lower throughput compared to the baseline.
Surprisingly, hybrid RotorNet also lags behind direct-circuit routing.
We attribute this gap to the TCP performance under packet reordering, as measured in Fig.~\ref{fig:transport}(b).
To verify our observation, 
we increase the TCP dupack threshold from the default 3 to 5, which almost eliminates reordering events in hybrid RotorNet and pushes the throughput back to the expected value, i.e., near \uGbps{25} with 50\% of data through the electrical fabric and 50\% of data through the optical fabric.
While slightly improved, VLB throughput is still low due to excessive reordering.

This case study demonstrated the process of troubleshooting transport performance, and it shows how researchers can tune transport protocol parameters and evaluate newly designed protocols for optical \dcns with \name.

\para{Case III: choice of optical hardware.}
The optics community has developed numerous \ocses, focusing on device-level characteristics like port count, reconfiguration delay, loss, and cost. Network architects traditionally select devices based on a general understanding of trade-offs between these factors, without insight into how these attributes impact network performance.
In this case study, we demonstrate how \name facilitates more informed decisions on optical hardware selection through its emulation capabilities.

\begin{figure}[t]
    \centering
    \begin{subfigure}[t]{0.49\columnwidth} \centering 
        \includegraphics[height=3.05cm]{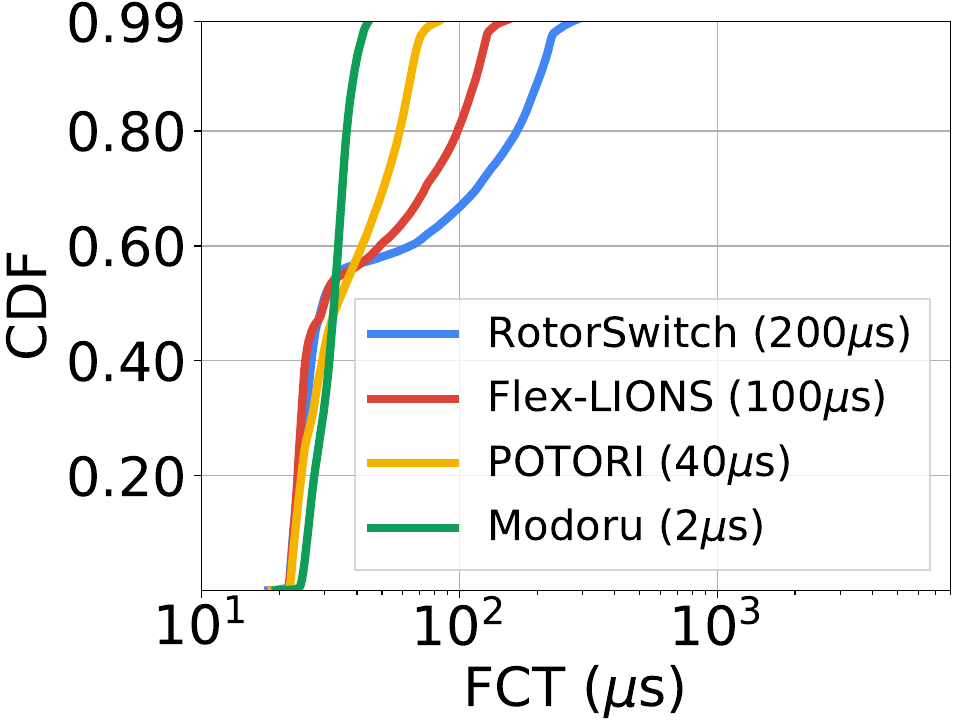}
        \phantomcaption\label{fig:ocs-vlb}
    \end{subfigure}
    \begin{subfigure}[t]{0.49\columnwidth} \centering 
        \includegraphics[height=3.05cm]{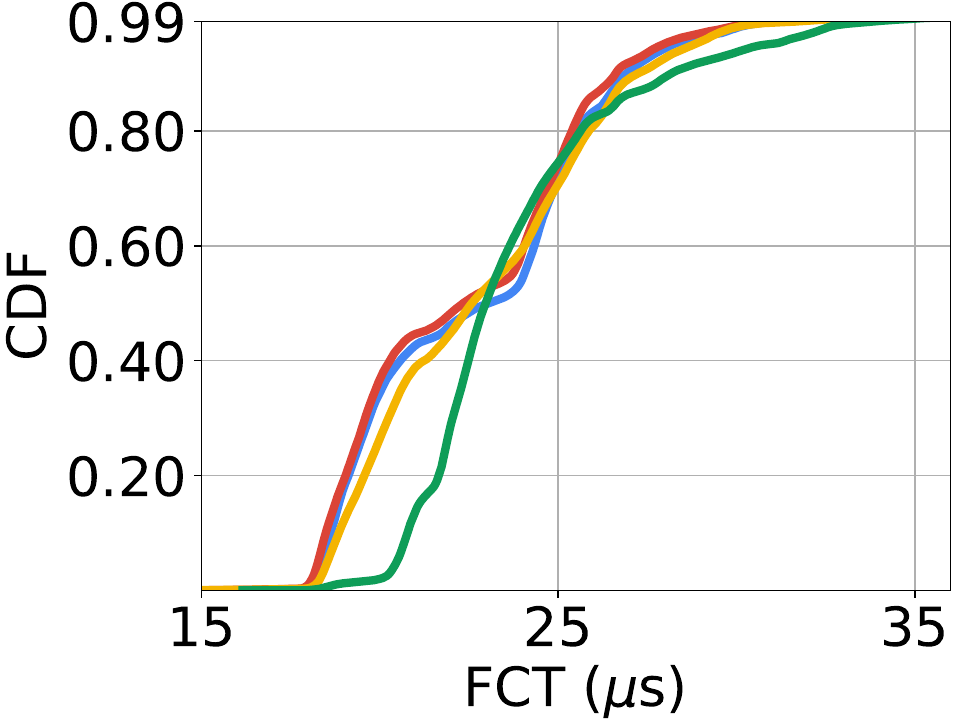}
        \phantomcaption\label{fig:ocs-ucmp}
    \end{subfigure}

    \figcap{Mice flow \fcts on RotorNet with \ocses of different time slice durations, under (a)VLB and (b)UCMP routing.}\label{fig:ocs-rotornet}
\end{figure}

We sample four recently proposed \ocs technologies and emulate the RotorNet architecture with them by inputting their physical characteristics and \ocs structures into the static configuration file. Fig.\ref{fig:ocs-rotornet} shows the \fcts for the Memcached application (as in Fig.~\ref{fig:memcached}) relative to the supported time slice duration of each \ocs device.

In Fig.~\ref{fig:ocs-vlb}, RotorNet's native VLB routing exhibits long tail \fcts proportional to the time slice duration, as packets are randomly routed through intermediate ToRs. In the worst case, a packet waits an entire optical cycle at the intermediate ToR for a direct circuit to the destination. This result suggests the shorter the time slice, the merrier, but \ocs costs rise substantially with shorter time slices. Fortunately, Fig.~\ref{fig:ocs-ucmp} presents a different trade-off with UCMP routing, which strategically selects intermediate ToRs, reducing \fcts and making performance less sensitive to time slice duration. UCMP is less effective under very short time slices, such as \umus{2}, due to higher risks of missing the planned time slice and being deferred to later slices. The best performance occurs at \umus{100} time slices, with little difference at \umus{200}, allowing good performance with more cost-effective \ocs devices.

We have demonstrated finding the performance-cost sweet spot for \ocs devices with a specific workload. \name enables deeper emulations for researchers and network architects to explore the interplay between optical hardware, network architectures, routing, and higher-layer protocols.

\section{Performance Benchmarks}\label{sec:perf}

We now evaluate the system components at scale and reproduce results from existing architecture implementations. We create a large-scale setting with the testbed equipment in Fig.~\ref{fig:testbed} and run \dcn traffic traces.
\name users can adopt this approach for scalable analysis with limited hardware.

\para{Experimental setup.}
We emulate the 108-ToR topology from the Opera paper~\cite{Opera} because its rigid ToR and \ocs structures make it compatible with other optical architectures. In Opera, each ToR has six \uGbps{100} uplinks to the optical network fabric and six \uGbps{100} downlinks to hosts. We implement a single representative ToR on a Tofino2 switch, referred to as the \textit{observed ToR}, and connect it to another Tofino2 switch through six \uGbps{100} links, which emulates the optical network fabric.
We populate the full time-flow table on the observed ToR with entries for the 108-ToR network, and the emulated optical fabric operates at full network scale. Three servers with dual-port NICs are connected to the observed ToR, acting as six hosts with one \uGbps{100} link each.
We replay the widely-used RPC~\cite{HOMA}, Hadoop~\cite{roy2015inside}, and KV store~\cite{atikoglu2012workload} \dcn traces on the hosts and scale the load to reach 40\% core link utilization as in production \dcns~\cite{FB-trace}.

\begin{figure}[tbp]
    \centering
    \begin{minipage}[t]{0.48\columnwidth} \centering
    \includegraphics[width=\linewidth]{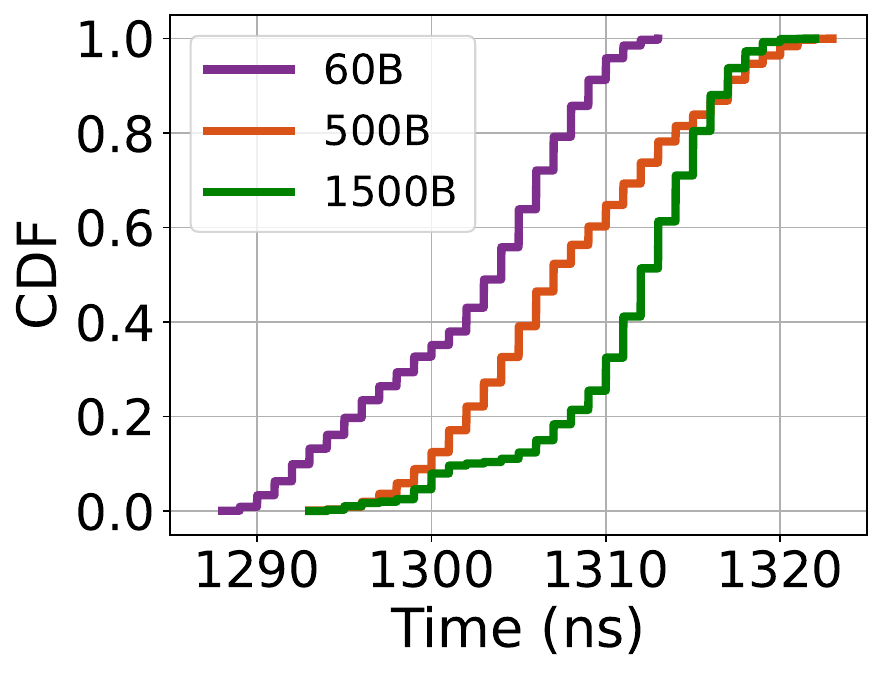}%
    \figcap{switch-to-switch delay with diff. packet sizes.}\label{fig:rotation-offset}
    \end{minipage}
    \hfil %
    \begin{minipage}[t]{0.48\columnwidth} \centering
    \includegraphics[width=\linewidth]{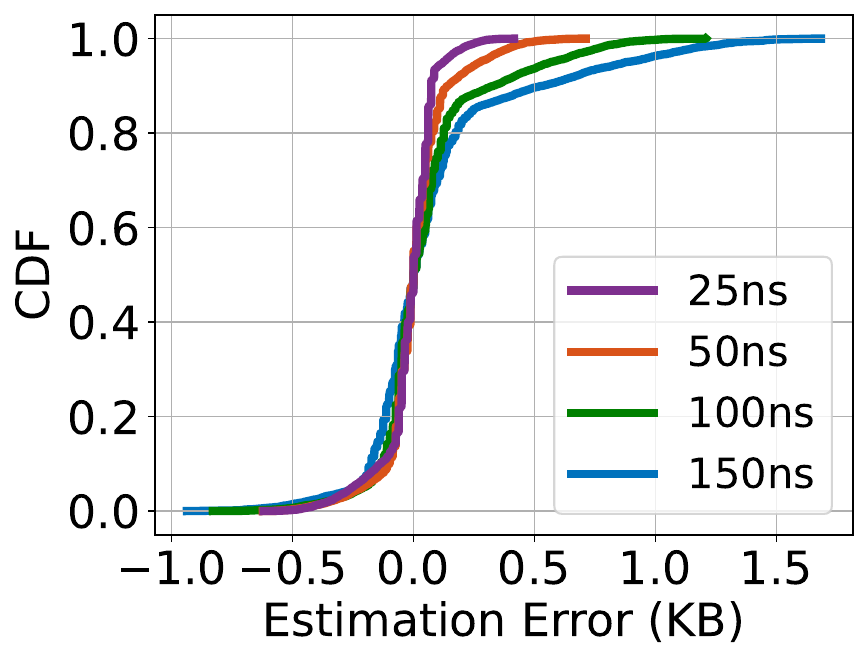}
    \figcap{EQO error under diff. update intervals.}\label{fig:es_error}
    \end{minipage}
\end{figure}

\para{Efficiency of queue management system.}
Rotation of calendar queues (\S\ref{sec:q-rotation}) must be aligned precisely with the time slices, but delays between switches, including the switch pipeline processing, packet serialization, and on-wire propagation delays, interfere with the accuracy of time calculation. 
We measure these delays from the observed ToR back to itself through the MEMS \ocs to use the same clock for accurate timestamping. This is done by continuously sending packets at line rate using the on-chip packet generator on the switch. The ToR-to-ToR delay is from when queue rotation is triggered on the sender side to when each packet arrives at the Rx MAC of the receiver side. Fig.~\ref{fig:rotation-offset} plots the delays with different packet sizes. The minimum delay is $\uns{1287}$, which we can offset by starting queue rotation earlier to ensure that the least delayed packet catches the upcoming circuit. The maximum delay is $\uns{1324}$, indicating that the most delayed packet arrives $1324-1287=\uns{34}$ after the circuit is established.
No packet should be sent in this \uns{34} to avoid packet loss. As will be discussed soon, this only wastes $\uns{34} / \umus{2} = 1.7\%$ of the \umus{2} minimum time slice duration.

\para{Queue occupancy estimation accuracy.}
The error of queue occupancy estimation (\S\ref{sec:infra_services}) is the difference between our estimated queue occupancy (EQO) from the ingress pipeline and the ground-truth queue occupancy read by an egress packet.
We measure it with the above setting (in Fig.~\ref{fig:rotation-offset}) but combine line-rate traffic and bursty traffic to fill and drain the queue periodically.
The estimation accuracy increases with the update interval (Fig.~\ref{fig:es_error}).
We find \uns{50} draws a good balance between the estimation accuracy and the consumption of pipeline processing resources on the ToR. 
The estimation error is within \uB{725}, less than half MTU-size packet, 
and the packet generator sends one packet every \uns{50} (or at \uMpps{20}),
which creates only 1.3\% pipeline forward overhead on Tofino2 with \uBpps{1.5} processing capacity.

\para{Minimum time slice duration.}
We derive the minimum achievable time slice duration in \name to evaluate its generality to support various optical proposals. To ensure a duty cycle above 90\%, the slice duration is typically set to at least 10$\times$ the guardband, which covers the maximum of circuit reconfiguration and unavoidable system delays, preventing data loss. This analysis focuses on system limits, assuming \ocs hardware is not a limiting factor, and asks what time slice duration \name can achieve in that case.

The \uns{34} queue rotation variance (Fig.~\ref{fig:rotation-offset}) should be covered by the guardband. 
In addition, the queue occupancy estimation error of \uB{725} (Fig.~\ref{fig:es_error}) translates to \uns{58} delay under \uGbps{100} bandwidth,
meaning packets may get impacted by false negatives for \uns{58}, where a full queue is mistakenly seen as not full. Besides, our synchronization work shows up to \uns{28} sync errors in a 192-ToR optical \dcn~\cite{OpSync}, which requires a guardband of $28 \times 2 = \uns{56}$ for clock discrepancies above and below the actual clock. Therefore, the total guardband is $34 + 58 + 56 = \uns{148}$, and with added headroom for runtime variations, we set the guardband to $\uns{200}$. \name thus supports a minimum time slice duration of $\uns{200} \times 10 = \umus{2}$, the shortest time slice achievable with commodity devices.
We observe no packet loss in all the experiments with this guardband value.

\begin{table}[t]
    \begin{minipage}{0.46\columnwidth} \centering
        \includegraphics[height=3.0cm]{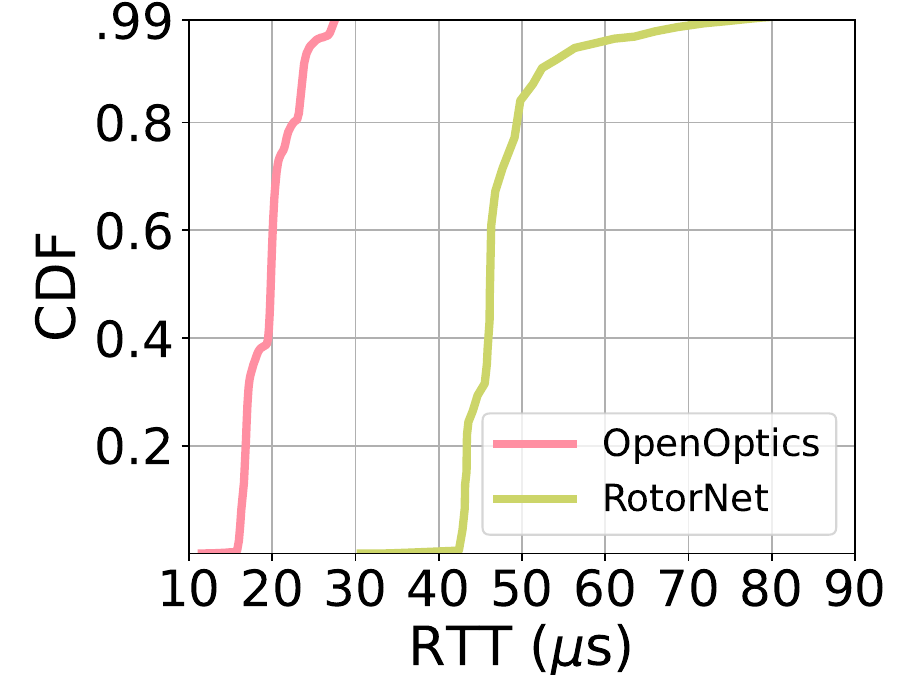}\label{fig:comp_openoptics}
        \captionof{figure}{\textnormal{\it UDP latency in \name
        vs. 
         in RotorNet
         (Fig.15 blue curve in~\cite{RealizeRotorNet}).}}\label{fig:reproduce}
    \end{minipage}
    \hfill
    \begin{minipage}{0.46\columnwidth} \centering
        \footnotesize
        \tabcap{Resource usage of \name on Tofino2.}\label{tb:resource}
        \begin{tabular}{lr}
        \toprule
        \thead{Resource} & \thead{Usage} \\ \midrule
        SRAM         & $3.8\%$  \\
        TCAM         & $2.3\%$\\
        Stateful ALU & $9.4\%$  \\
        Ternary Xbar & $13.8\%$\\
        VLIW Actions & $5.6\%$  \\
        Exact Xbar   & $7.8\%$\\
        \bottomrule
        \end{tabular}
    \end{minipage}
\end{table}

\para{Emulation accuracy.}
We verified \name{}'s correctness in Fig.~\ref{fig:applications} on both real and emulated fabrics.
We now augment this validation by reproducing experimental results from ``Realizing RotorNet''~\cite{RealizeRotorNet}, the only \tro architecture implemented so far, with actual \ocses.
We replicate the UDP RTT latency experiment by continuously sending UDP packets from one host to another and measuring the RTT for each packet.
Per Fig.~\ref{fig:reproduce}, \name'{} results on the emulated optical fabric and RotorNet's on real \ocses exhibit similar latency distribution patterns, with stepped RTT increases corresponding to additional routing hops.
The similar curve shapes indicate comparable routing behaviors between our emulated fabric and the actual \ocs fabric, confirming correctness of \name.
\name achieves lower RTTs and eliminates the long tail because our switch system and \vma stack produce less delay and variance than RotorNet's FPGA NICs and the kernel UDP implementation.

\para{Switch resource usage.}
As listed in Table~\ref{tb:resource}, the resource usage of an \name-enabled ToR in the 108-ToR \dcn is a small percentage.
The low usage of SRAM, VLIW Actions, and TCAM indicates the efficient implementation of registers and lookup tables. 
Stateful ALU and Ternary Xbar have higher usage, due to the arithmetic calculations and branching operations for slice-miss detection.
All the resources are under 13.8\% of an Intel Tofino2 switch's capacity,
leaving sufficient room for \name to scale up to larger \dcns.

\para{Optimization efficiency.}
We evaluate \name's optimization options in Appx.~\ref{sec:eval-detail} using architectures and routing schemes that require them. Buffer offloading maintains switch buffer usage well below Tofino2's limit, even for buffer-intensive VLB routing, with offloaded packets reloaded to switches quickly and predictably. Congestion detection, paired with traffic pushback, facilitates and enhances existing congestion reactions, ensuring high throughput and zero loss. We also break down the performance gains of these two features to demonstrate their individual effectiveness.

\section{Related Work}\label{sec:related_work}%

In \S\ref{sec:background}, we presented an overview of \dcn architectures~\cite{Helios, cThrough, OSA, Mordia, Quartz, MegaSwitch, WaveCube, ProjectToR, Flat-tree, ShareBackup, OmniSwitch, RotorNet, Sirius, Opera, SiP-ML, TopoOpt, JupiterEvolving, Shale, NegotiaToR}, and implemented six of them
over the \umus{2} minimum time slice duration (\S\ref{sec:research}), showing generality of \name. 
Our time-flow table abstraction applies to all architectures, regardless of constraints from today's commodity devices.

UCMP~\cite{UCMP} and HOHO~\cite{HOHO} are general routing algorithms for \tro architectures. We realized them on \name and found out UCMP could reduce RotorNet's performance sensitivity to optical hardware, potentially allowing for the use of more cost-effective \ocses (\S\ref{sec:research}).
Recent transport proposals of reTCP~\cite{reTCP} and TDTCP~\cite{TDTCP} highlight the need for system designs atop of optical architectures, which validates the design purpose of \name. Our TCP throughput case study (\S\ref{sec:research}) confirms packet reordering addressed in these works.

Etalon~\cite{reTCP} and ExReC~\cite{ExRec} are the closest evaluation platforms for optical \dcns, but they emulate switches on hosts without a full switch stack, also achieving limited scale (8 ToRs) and capacity (\uGbps{10})~\cite{reTCP, ExRec}. In contrast, \name is more realistic with the full switch and host implementations. We have demonstrated feasibility of \name in a 108-ToR setting at $\uGbps{100}$ (\S\ref{sec:perf}), and it can support higher capacity with better equipment. Additionally, \name offers a comprehensive solution, including real \ocs support, emulated \ocses, and Mininet.

\section{Conclusion}\label{sec:conclusion}

We presented the design and realization of \name, an open research framework that promotes system innovation for optical \dcns.
Much of the prior work on optical \dcns{} focus on the optics part, while relegating the systems side to high-level simulations.
\name addresses this status quo:
\name decouples the software system from the optical architecture.
We present meticulously designed abstractions (e.g., time-flow table), a simple and powerful programming model, and intuitive services that serve as building blocks to realize complex designs.
The result is a research framework that lowers the barrier to investigate and explore optical \dcns{}.
\name{} allows researchers to port the immense knowledge we have accrued from experimenting with traditional \dcns{} carefully and systematically to optical \dcns{}.

\bibliographystyle{plain}
\bibliography{ref}

\clearpage
\begin{appendices}

\section{Queue Occupancy Estimation}\label{sec:slice-miss}

We implement queue occupancy estimation using a register array in the ingress pipeline to track each calendar queue's occupancy. Ideally, these occupancy registers should be updated whenever packets are enqueued and dequeued. However, since registers at the ingress pipeline can only be updated by ingress packets, we increment the queue's occupancy by the packet size upon enqueueing, while for dequeuing, we estimate occupancy reduction through periodic updates.

We use packet generator to create an ingress packet at every \textit{update interval} to trigger updates. As dequeuing happens at line rate, the packet triggers the occupancy register of the active queue to decrease by the product of the \textit{link bandwidth} and the \textit{update interval}, or sets it to zero if the result is negative, indicating the queue has emptied.
As shown in Fig.~\ref{fig:es_error}, an update interval of $\uns{50}$ results in less than one packet estimation error at minimal pipeline processing overhead.

We make queue occupancy estimation a common module for slice-miss prevention in \tro architectures. For example, Sirius and Shale query the occupancy of candidate intermediate nodes to send packets to the least occupied ones~\cite{Sirius, Shale}; Opera caps queue occupancy and drops packets exceeding this limit~\cite{Opera}; UCMP and HOHO check queue occupancy to defer packets that cannot fit within the intended time slice to a later one~\cite{UCMP, HOHO}.
Besides supporting these mechanisms, we also use push-back messages (\S\ref{sec:infra_services}) to halt flows at their origin, preventing overwhelming the circuits.

\begin{table}[bp]
\vspace{5pt}
\footnotesize
\centering
\tabcap{99.9\%-ile buffer usage under different traces with \umus{300} slice duration. Total buffer on Tofino2 is \uMB{64}.}
\begin{tabular}{llll}
\hline
\thead{Routings} & \thead{VLB (offloaded)}   & \thead{HOHO}   & \thead{UCMP}              \\ \hline
KV Store         & \uMB{9.48} (\uMB{1.26})      & \uMB{2.40}     & \uMB{2.44}       \\
RPC              & \uMB{9.96} (\uMB{1.38})      & \uMB{3.06}     & \uMB{4.15}       \\
Hadoop           & \uMB{12.78} (\uMB{1.56})     & \uMB{3.90}     & \uMB{6.54}       \\ \hline
\end{tabular}\label{tab:buffer}
\end{table}

\para{Buffer offloading efficiency.}
We have shown good performance of the host system in Fig.~\ref{fig:transport}: flow pausing (\S\ref{sec:infra_services}) ensures expected throughput (50\% of maximum throughput with 50\% circuit availability) for direct-circuit routing without packet reordering. Here, we evaluate the stability of RTTs between switches and hosts, which is critical for buffer offloading. 

\begin{figure}[t] \centering
    \includegraphics[height=3.0cm]{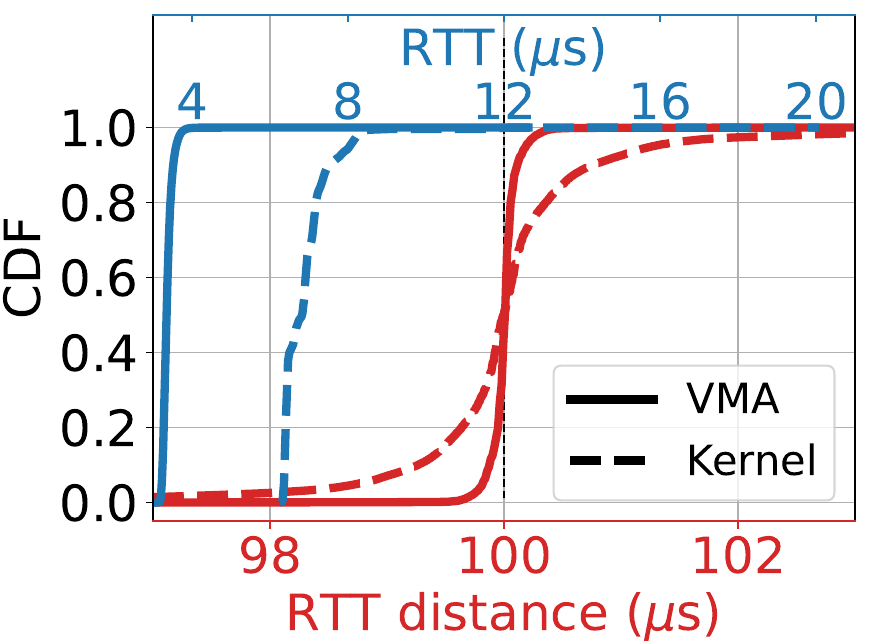}%
    \figcap{RTT delay (blue) and RTT distance to the \umus{100} interval (red).}\label{fig:vma}
\end{figure}

We send \uB{1500} packets from the observed ToR to a connected host at \umus{100} intervals. The host returns them upon receipt to simulate offloading and retrieval. Fig.~\ref{fig:vma} shows our \vma-based implementation ensures stable RTTs: 95\% of RTTs exhibit a small variance of \umus{0.75}, and their deviation from the expected \umus{100} intervals remains within $\pm$ \umus{0.25}. The performance is significantly better compared to a kernel module baseline, confirming the efficiency of \vma.
In practice, we return offloaded packets to the switch early to offset the base delay and variance.
Packets arriving marginally earlier are buffered on the switch consuming minimal memory.

\para{Switch buffer usage.}
VLB, HOHO, and UCMP are the only routing solutions that have packets wait at intermediate nodes, so we evaluate switch buffer usage with those. 
We run \umus{300} time slices, considered long for \tro architectures. As shown in Table~\ref{tab:buffer}, HOHO and UCMP maintain low buffer usage across traces, as they prioritize latency by directing packets to the nearest time slices. In contrast, VLB causes packets to wait for extended periods at intermediate nodes.
Nonetheless, the maximum buffer usage of \uMB{12.78} remains well below the \uMB{64} limit of Tofino2 switches, and buffer offloading to hosts (\S\ref{sec:infra_services}) reduces the buffer load to a minimal level.
These results indicate that the calendar queues (\S\ref{sec:q-rotation}) in our switch system are sustainable for higher bandwidth in the future.

\section{Extra Benchmarking Results}\label{sec:eval-detail}

\begin{table}[bp]
\footnotesize
\vspace{5pt}
\tabcap{{Effectiveness of congestion detection and traffic push-back in HOHO with Hadoop/RPC/KV-store traces under \umus{300} time slices.}}
\label{tb:ac-pb}
\centering
\begin{tabular}{cccc}
\toprule
\thead{Congestion Detection}      & \xmark          & \cmark      & \cmark   \\ 
\thead{Traffic Pushback}        & \xmark          & \xmark          & \cmark   \\ \hline
Throughput (Gbps) & 67/69/64          & 67/69/64          & 50/54/49       \\
Loss Rate        & 1.1\%/1.0\%/2.1\% & 1.1\%/1.0\%/2.0\% & 0\%/0\%/0\%    \\
Average Delay   & 160/150/\umus{98}     & 125/144/\umus{91}     & 6/5/\umus{8} \\
95\%-ile Delay   & 2.2/2.2/\ums{2.1}     & 2.1/2.2/\ums{1.9}     & 85/90/\umus{83} \\

\bottomrule
\end{tabular}
\end{table}

\para{Effectiveness of congestion detection and traffic push-back.}
Among the routing schemes, HOHO is most vulnerable to congestion. It minimizes latency by always sending packets over the earliest available time slices, which can lead to overshooting and overwhelming the slice. When congestion is detected, HOHO defers packets that cannot fit into the scheduled time slice to a later one, but it does not implement congestion control to stop or slow down incoming traffic. In this case, traffic push-back can be enabled as a last line of defense when congestion persists. Therefore, we evaluate the effectiveness of our congestion detection and traffic push-back mechanisms (\S\ref{sec:infra_services}), specifically with HOHO.

We stress-test the calendar queues by increasing traffic load to 70\% core link utilization, triggering the congestion detection and traffic push-back mechanisms. As shown in Table~\ref{tb:ac-pb}, 
without congestion detection or push-back (column 1), packets are always enqueued to the most desirable calendar queue, even when it exceeds the slice capacity. The queue pauses after the time slice ends, holding buffered packets until the next cycle(s), leading to long delays and packet loss. With congestion detection alone (column 2), loss rates and average delay decrease slightly as packets are deferred to later time slices when the primary queue is full, but queues eventually fill up, causing losses. When both mechanisms are combined (column 3), push-back engages once the primary queue is full, and slice-miss detection handles in-flight traffic before senders react. This eliminates packet loss, and low queuing delay indicates most packets are enqueued in the primary calendar queue.
These measures have similar effects on Opera and UCMP, reducing packet loss rate from 2.5\% to 0\% in Opera, also eliminating loss and reducing tail latency from \ums{1.4} to \umus{115} in UCMP.

\end{appendices}

\end{document}